\documentclass[12pt]{article}

\usepackage{amsmath}
\usepackage{amssymb}
\usepackage{amsfonts}
\usepackage{mathtools} 

\usepackage{url}
\usepackage{xcolor} 
\usepackage{graphicx} 
\usepackage{float} 
\usepackage{booktabs}
\usepackage{tabularx}
\usepackage{subcaption} 

\usepackage[a4paper, left=1in, right=1in, top=1in, bottom=1in]{geometry}

\usepackage{setspace}
\usepackage[round]{natbib}
\usepackage{authblk} 

\usepackage[acronym]{glossaries}
\newacronym{gps}{GPS}{Global Positioning System}
\newacronym{dmr}{DMR}{Dirichlet-multinomial regression}
\newacronym{dm}{DM}{Dirichlet-multinomial}
\newacronym{pca}{PCA}{Principal Component Analysis}
\newacronym{pcs}{PCs}{principal components}
\newacronym{pc1}{PC1}{first principal component}
\newacronym{pc2}{PC2}{second principal component}
\newacronym{pc3}{PC3}{third principal component}
\newacronym{pc4}{PC4}{fourth principal component}
\newacronym{irb}{IRB}{Institutional Review Board}
\newacronym{fwer}{FWER}{family-wise error rate}
\newacronym{imus}{IMUs}{inertial measurement units}

\usepackage{tikz}
\usetikzlibrary{shapes.geometric, arrows.meta, positioning}
\tikzstyle{startstop} = [rectangle, rounded corners, minimum width=3cm, minimum height=1cm,text centered, draw=black]
\tikzstyle{process2} = [rectangle, minimum width=3cm, minimum height=1cm, text centered, draw=black, fill=red!10]
\tikzstyle{processblue} = [rectangle, minimum width=3cm, minimum height=1cm, text centered, draw=black, fill=blue!10]
\tikzstyle{arrow2} = [thick,->,>=stealth]

\usepackage{titlesec}
\titleformat{\section}[block]{\normalfont\large\bfseries}{\thesection}{1em}{}
\titleformat{\subsection}[block]{\normalfont\bfseries}{\thesubsection}{1em}{}
\titleformat{\paragraph}[block]{\normalfont\bfseries}{\theparagraph}{1em}{}

\title{Movement Dynamics in Elite Female Soccer Athletes: The Quantile Cube Approach}
\author[1]{Kendall L. Thomas}
\author[1]{Jan Hannig}

\affil[1]{Department of Statistics and Operations Research, University of North Carolina at Chapel Hill, Chapel Hill, NC, US} 
\date{}

\begin{document}

\maketitle

\begin{abstract}
\label{sec:abstract}
This paper presents the \emph{quantile cube}, a novel three-dimensional summary representation designed to analyze external load using GPS-derived movement data. While broadly applicable, we demonstrate its utility through an application to data from elite female soccer athletes across 23 matches. The quantile cube segments athlete movements into discrete quantiles of velocity, acceleration, and movement angle across match halves, providing a structured and interpretable framework to capture complex movement dynamics. Statistical analysis revealed significant differences in movement distributions between the first and second halves for individual athletes across all matches. Principal Component Analysis identified matches with unique movement dynamics, particularly at the start and end of the season. Dirichlet-multinomial regression further explored how factors such as athlete position, playing time, and match characteristics influenced movement profiles. Our analysis reveals external load variations over time and provides insights into performance optimization. The integration of these statistical techniques demonstrates the potential of data-driven strategies to enhance athlete monitoring and workload management in women's soccer.
\end{abstract}


\section{Introduction} 
\label{sec:intro}

Wearable technology has fundamentally transformed how athletic performance is monitored and analyzed, especially among elite athletes. These devices generate vast quantities of data, providing insights into the physical demands placed on athletes and enabling more precise adjustments to training regimens \citep{cummins_2013}. Building on this foundation, integration with advanced data analytics has become essential for extracting actionable insights, especially when evaluating training volume and intensity \citep{bourdon_2017}. 

A key component of this process is external load monitoring via \gls{gps} technology. These systems quantify movement metrics such as velocity, acceleration, and distance covered during matches and training. Such metrics capture critical aspects of movement intensity and dynamics, allowing for the assessment of workload distribution, fatigue development, and positional movement demands. While total distance and average speed are often reported, more nuanced metrics such as acceleration and deceleration efforts and individualized speed thresholds may provide more sensitive indicators of player workload and fatigue \citep{Snyder2024}.

Despite the potential of these metrics, challenges remain. Many wearable device companies provide proprietary ``training load" metrics that integrate multiple performance variables without disclosing their exact formulas or component definitions. For example, speed threshold definitions, which categorize movement into zones such as high-speed running or sprinting, can vary across devices, teams, or sports. Some use absolute velocities (e.g., $>5$ m/s), while others reference an athlete’s maximum speed (e.g., $>70\%$ or $>90\%$). Consequently, comparisons across devices can be difficult, leading to inconsistencies in training and performance interpretation, and thereby raising questions about metric reliability.

While most of the research on workload patterns has largely focused on male athletes, studies investigating women's soccer are steadily increasing. \citet{DeLucia2024} reported gender differences in GPS-derived workload metrics such as sprint distance, accelerations, and player load per minute. Extending this work, \citet{Kuhlman2025} highlighted sport- and position-specific workload variations across women’s collegiate soccer, lacrosse, and field hockey. Within women's soccer, traditional \gls{gps} metrics do not always correlate strongly with match outcomes \citep{Gailor2024}. Time-segmented analyses reveal early-onset fatigue, which affects high-speed running, acceleration, and deceleration during match play \citep{Snyder2024}. This highlights short-term performance declines that composite metrics over a full session, such as total load, can obscure. Collectively, these findings emphasize the need for refined, interpretable, and position-specific monitoring frameworks tailored to female athletes.

Temporal and positional variations in workload demands have been well documented in male professional players. For example, \citet{barrera_2021} observed reductions in high-speed running and other external load metrics during the second half of professional matches, reflecting fatigue and tactical adjustments. Similarly, \citet{wehbe_2014} reported that midfielders cover greater total and high-intensity running distances than defenders, highlighting position-specific load profiles. While these studies focus on male athletes, they provide a valuable comparative framework for investigating similar dynamics in elite female soccer athletes using advanced modeling techniques.

Modern statistical approaches have emerged to address the computational challenges of analyzing large, longitudinal \gls{gps} datasets. Traditional analyses often assume independent and identically distributed data, which rarely holds in real-world contexts \citep{luo_2020}. Recent methods, including linear state-space mixed models \citep{luo_song_2023} and incremental inference via dynamic updates \citep{luo_wang_hector_2023}, leverage the summation of summary statistics over data batches to dynamically update point estimates and standard errors. However, reliance on summary statistics can obscure important extremes of the data distribution, which are critical for capturing nuanced patterns and generating actionable insights. To overcome these limitations, researchers are increasingly adopting sophisticated approaches that integrate multiple data sources for a more comprehensive understanding of longitudinal workloads.

Complementing these statistical advances, the integration of GPS-derived movement metrics with multi-dimensional and machine learning frameworks has shown considerable promise for improving workload monitoring and injury prediction in male athletes \citep{vallance_2020, rossi_2018}. These approaches leverage both external load data (e.g., velocity, acceleration, distance covered) and internal load indicators (e.g., subjective well-being, heart rate) to create richer, more predictive models of athlete performance and injury risk. However, the complexity of these models can limit their practical application for coaching and training staff, as interpretability and real-time usability are often constrained. \citet{ferraz_2023} emphasize the urgent need for integrative frameworks that combine external and internal load data in a manner that is both statistically robust and practically actionable. Moreover, the implementation and validation of such methods in women’s sports remains limited, leaving a critical gap in evidence-based monitoring strategies for female athletes. Developing interpretable, multi-dimensional models tailored to the unique physiological, tactical, and positional demands of female soccer athletes represents a key step toward bridging this gap and translating advanced analytics into meaningful coaching and training interventions.

To address these gaps, we propose a novel method to integrate \gls{gps}-based external load metrics with athlete and match characteristics in elite female soccer. Our primary objective is to develop interpretable statistical models that quantify movement patterns---specifically velocity, acceleration, and movement angle---and examine their relationships with athlete performance and match outcomes. Unlike traditional zone-based thresholds, which rely on arbitrary cutoffs and vary across devices, our approach leverages the full empirical distribution of movement features to produce player-specific and statistically principled profiles. By combining probabilistic modeling with transparent statistical methods, this approach bridges the gap between data collection and practical application, providing a data-driven foundation for optimizing training protocols.

The paper is organized as follows: Section 2 details the data and proposed summaries for downstream analysis, Section 3 presents the methods and results, Section 4 discusses findings and practical implications, and Section 5 concludes with a summary of strengths, limitations, and directions for future research.

\section{Data}
\label{sec:data}

The data was collected by the Applied Physiology Lab in the Exercise Science Department at UNC Chapel Hill and shared under \gls{irb} 23-2673. A summary of the notation used throughout this section is provided in the Notation Table (see Appendix~\ref{appendix:notation}).

GPS tracking data were obtained from all match sessions over one season for 33 elite female soccer athletes. Only match sessions in which an athlete played at least 25 minutes in both the first and second halves were retained. Overtime periods were removed to ensure uniform match durations and comparability. Athletes who met this full-match play criterion in more than five match sessions were then selected. This filtering process resulted in a subset of nine athletes and 23 matches, yielding 198 valid athlete-match sessions. Note that not every selected athlete participated in every included match.

\begin{figure}[ht]
    \centering
    \includegraphics[width=0.45\textwidth, height=0.45\textwidth]{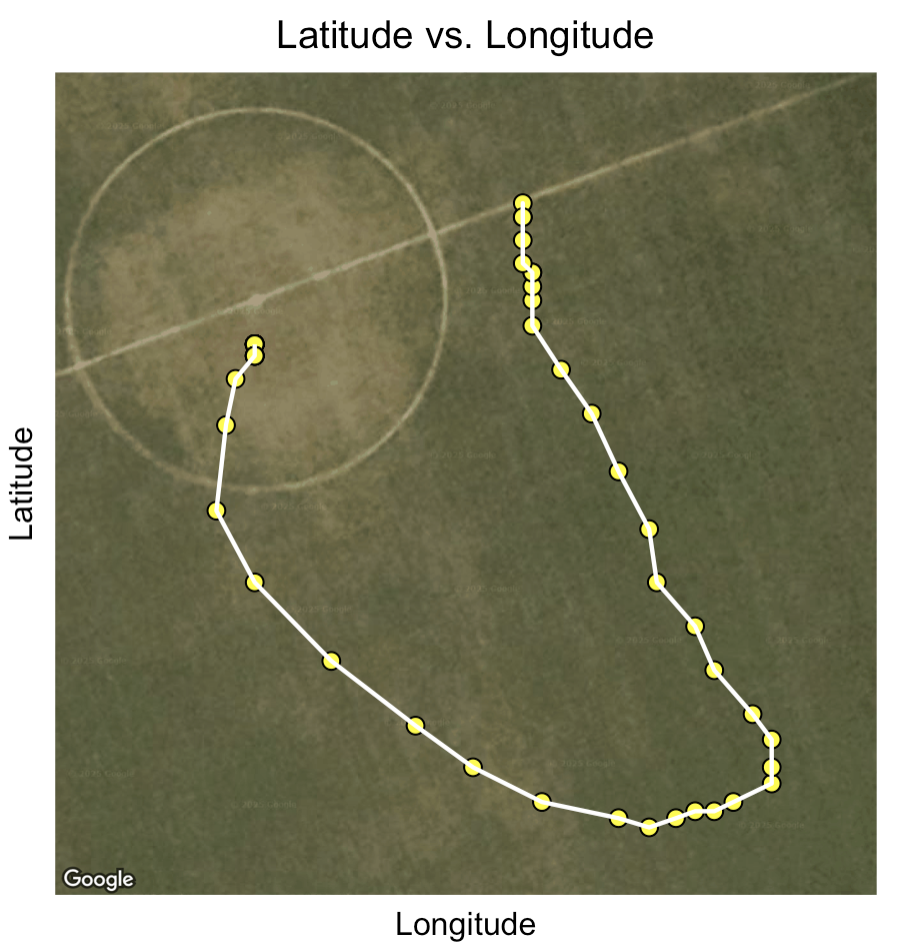}
    \includegraphics[width=0.45\textwidth, height=0.45\textwidth]{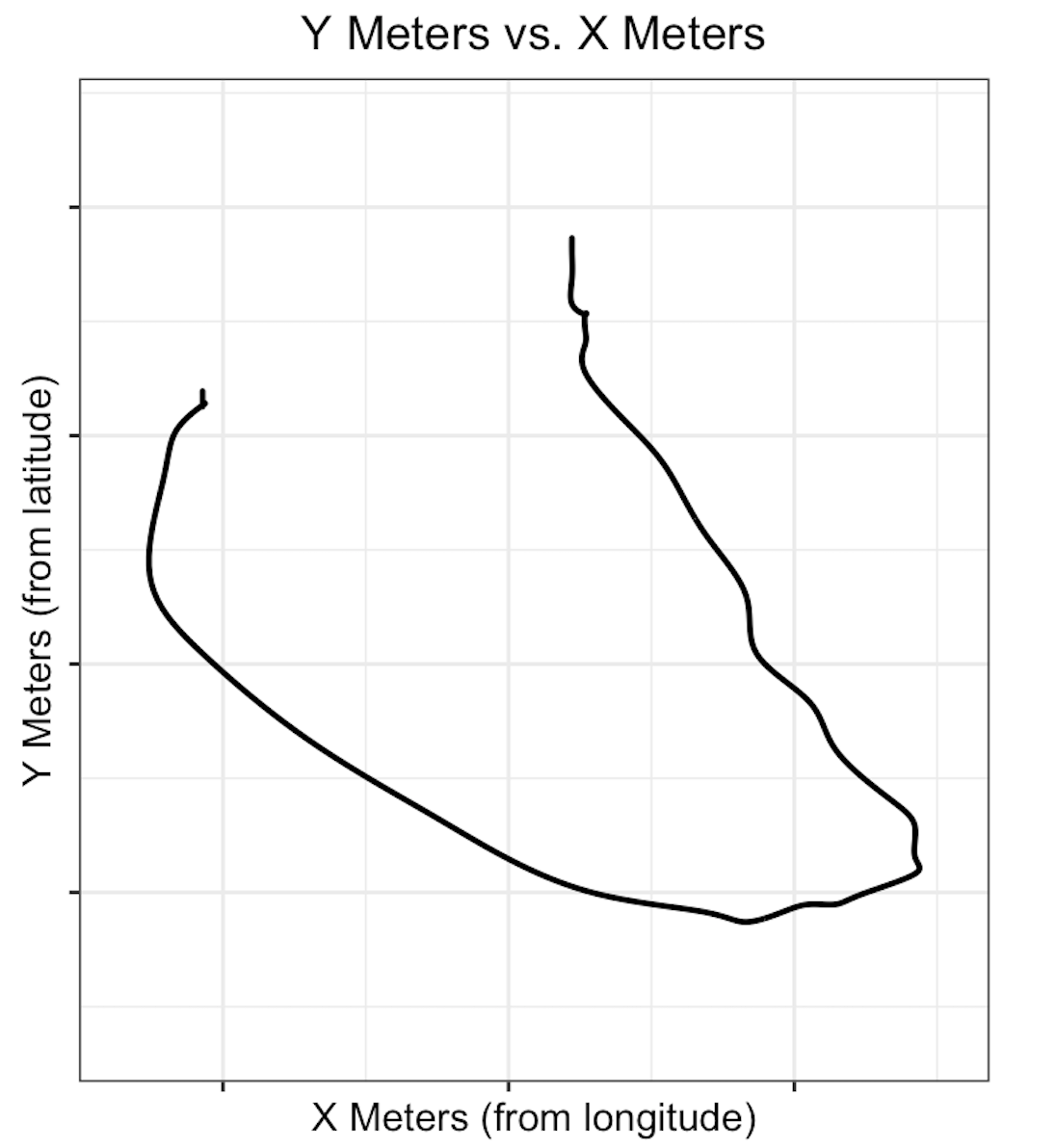}
    \caption{Left: Raw \gls{gps} data for 50 seconds of movement for one athlete overlaid on a satellite map (source: \citet{google_maps_api}). Right: Interpolating spline (in meters) fit to the same 50 seconds of movement for one athlete seen on the left. For confidentiality purposes, longitude, latitude, and transformed coordinates are not displayed.}
    \label{fig:rawdataspline}
\end{figure}

Each raw \gls{gps} dataset corresponded to a single athlete in a single match (i.e., one athlete-match session), and contained one data point per second, consisting of a timestamp along with longitude and latitude coordinates for the athlete’s location. For example, if an athlete played 80 minutes in a match session, the raw dataset comprising one athlete-match session would contain 4800 rows of timestamped positional data. Each athlete-match session contributed two halves to the analysis, resulting in a total of $n = 198 \times 2 = 396$ athlete-match-halves. Figure~\ref{fig:rawdataspline} (left) shows a 50-second example of this raw data overlaid on a satellite map \citep{google_maps_api}. 

To calculate velocity, acceleration, and angle of movement from the raw \gls{gps} coordinates for one athlete-match, the longitude and latitude values were converted to $(x, y)$ coordinates in meters using standard spatial transformations \citep{pebesma2018}. A third-degree interpolating spline was fitted to the data at ten points per second to model the athlete's movements (Figure~\ref{fig:rawdataspline} , right). Velocity (in m/s) and acceleration (in m/s\textsuperscript{2}) were derived from the first and second derivatives of the spline, respectively. The angle of movement was calculated as the angular difference between the velocity vector (direction of movement) and the acceleration vector (direction of change in velocity), capturing the degree of turning or directional change. The angle was computed modulo 360 and subsequently shifted to the range of -180 to 180 degrees for directional interpretability. To remove low-magnitude noise, velocity values below $0.01$ m/s and acceleration values below $0.001$ m/s\textsuperscript{2} were thresholded to zero. Due to the right-skewed nature of the raw distributions, $\log_{10}(1+\text{velocity})$ and $\log_{10}(1+\text{acceleration})$ transformations were applied for interpretability. From this point forward, the transformed values will be referred to simply as velocity and acceleration, except where specified in Table~\ref{tab:vel_acc_quantiles}.

\subsection{The Quantile Cube}
\label{sec:data_quantile_cube}

Raw wearable \gls{gps} data provide detailed, high-frequency measurements of athlete movement, including velocity, acceleration, and direction. However, direct analysis of these data is challenging due to noise, complexity, and variability across athletes and matches. To address this challenge, we introduce the quantile cube, a novel three-dimensional summary representation that discretizes key movement features into quantiles, capturing their joint distribution over time. Specifically, the quantile cube partitions velocity, acceleration, and movement angle into a structured grid of quantile bins along each dimension, forming a cube-shaped summary that represents how movement intensities and directions vary throughout a match. This approach enables a clear characterization of the time athletes spend in different types of movements and facilitates robust comparisons and trend detection within and across players and match contexts. To our knowledge, this is the first application of a quantile-based three-dimensional summary framework in sports movement analysis, providing a flexible and interpretable foundation for downstream statistical modeling and inference.

To form the quantile cube, the spline-derived data, containing velocity, acceleration, and angle of movement at ten points per second, was aggregated across all 396 athlete-match-halves. For velocity and acceleration, five bins  (0-20th, 20-40th, 40-60th, 60-80th, and 80-100th percentiles) were selected to provide a detailed characterization of movement intensity. The number of bins was selected to provide a compromise between a continuous representation of movement effort over time and having a large enough number of observations within each bin. The corresponding quantiles are shown in Table~\ref{tab:vel_acc_quantiles}.

\begin{table}[H]
\centering
\begin{tabular}{|c|c|c|c|c|c|c|}
\hline
\textbf{Quantile (\%)} & \textbf{0\%} & \textbf{20\%} & \textbf{40\%} & \textbf{60\%} & \textbf{80\%} \\
\hline
\textbf{Velocity (m/s)} & 0.0100 & 0.3289 & 0.9006 & 1.5026 & 2.5983 \\
\hline
\textbf{Acceleration (m/s\textsuperscript{2})} & 0.0000 & 0.4220 & 0.8159 & 1.2930 & 2.0367 \\
\hline
\end{tabular}
\caption{Five quantiles for velocity and acceleration (values shown are raw, prior to the log-transformation).}
\label{tab:vel_acc_quantiles}
\end{table}

\begin{figure}[H]
    \centering
    \includegraphics[scale=0.7]{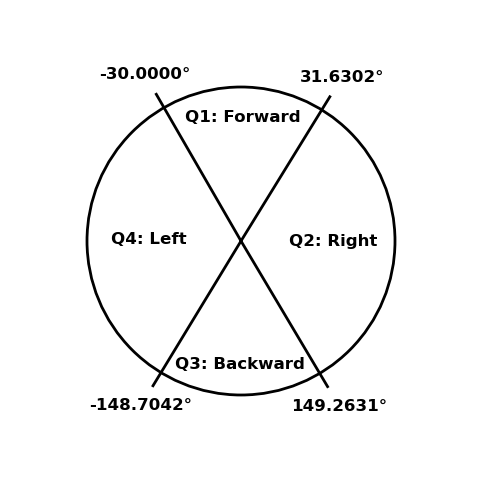}
    \caption{Segmentation of movement angles into four quantiles, starting from a shifted baseline of -30 to align with the four cardinal directions: forward, right, backward, and left.}
    \label{fig:angle_segmentation}
\end{figure}

For the angle of movement, four quantiles (0th, 25th, 50th, and 75th) were computed, starting from a shifted baseline of -30 degrees. This segmentation, illustrated in Figure~\ref{fig:angle_segmentation}, aligns with the four cardinal movement directions: forward, right, backward, and left. The numeric quantile cut-points for angle are reported in Table~\ref{tab:angle_quantiles}.

\begin{table}[H]
\centering
\begin{tabular}{|c|c|c|c|c|c|}
\hline
\textbf{Quantile (\%)} & \textbf{0\%} & \textbf{25\%} & \textbf{50\%} & \textbf{75\%} \\
\hline
\textbf{Angle (°)} & -30.0000 & 31.6302 & 149.2631 & -148.7042 \\
\hline
\end{tabular}
\caption{Four quantiles for movement angle, starting from a shifted baseline of -30°.}
\label{tab:angle_quantiles}
\end{table}

Conceptually, the quantile cube acts like a three-dimensional histogram that records how much time an athlete spends at combinations of velocity, acceleration, and movement angle. This provides an intuitive summary of movement style that can then be compared across halves, matches, or players.

Using the defined quantile boundaries derived from the full set of 396 athlete-match-halves, a quantile cube for each half of every athlete's match was constructed. Each athlete-match-half's spline-derived data were discretized using these fixed global boundaries, ensuring consistent binning across all sessions. Each dimension of the cube represents one of the key metrics: velocity, acceleration, and angle of movement. Color intensity within each cell indicates the proportion of time the athlete spent in that specific combination of velocity, acceleration, and angle quantiles.

\begin{figure}[ht]
    \centering
    \includegraphics[scale=0.6]{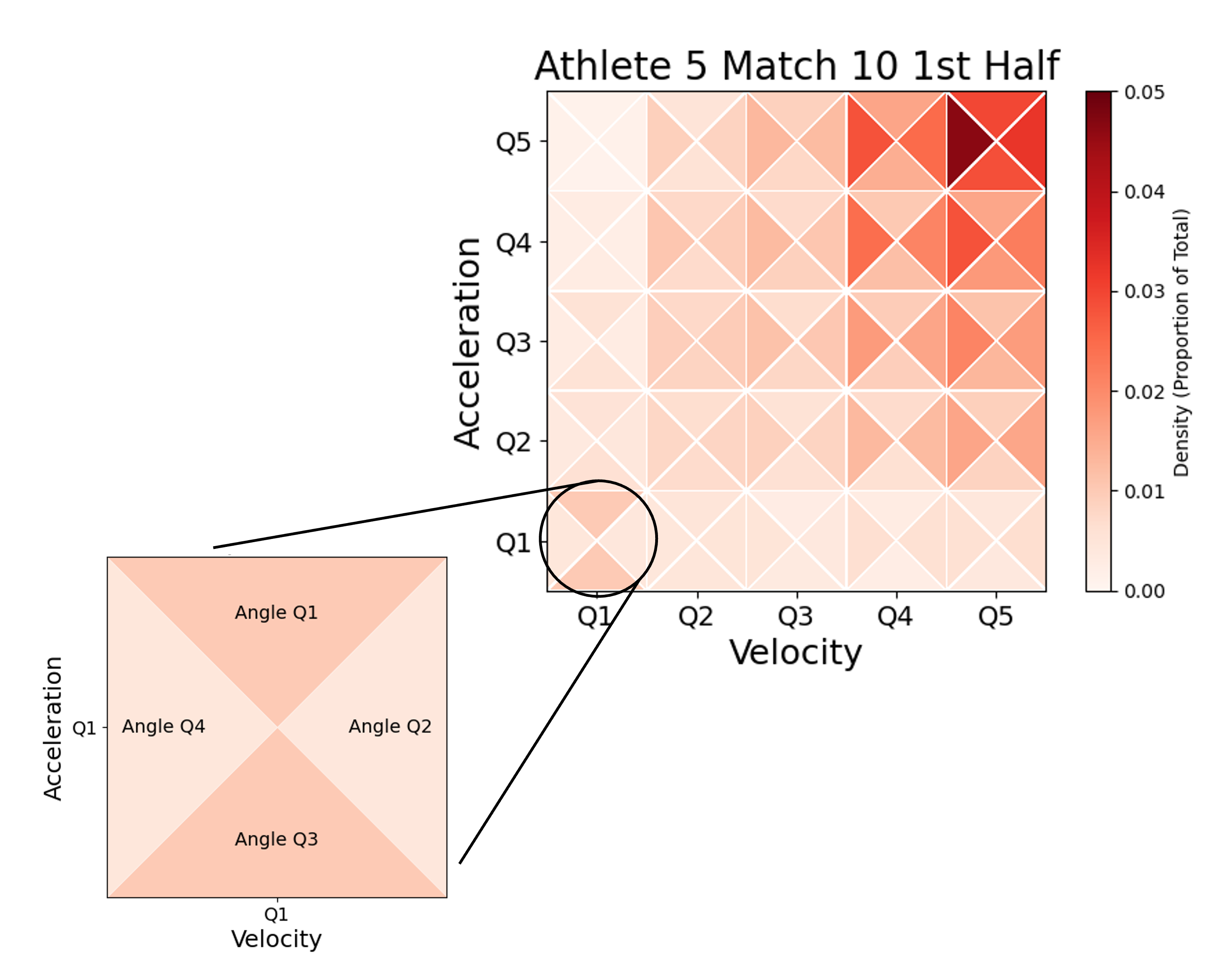}
    \caption{Visual representation of the quantile cube for the density of movements in the first half of Match 10 for Athlete 5. The main plot shows the distribution of movements across velocity (x-axis), acceleration (y-axis), and angle of movement quantiles, with color intensity indicating the proportion of time spent in each bin. The inset zooms into the first quantile for velocity and acceleration, illustrating the subdivision of the four angle quantiles.}
    \label{fig:quantileCube}
\end{figure}

The quantile cube can be visualized as shown in Figure~\ref{fig:quantileCube}, where the inset zooms into a single velocity-acceleration bin to illustrate how the four angle quantiles are further subdivided within that bin. For example, this figure shows that the largest proportion of time (0.04629) was spent in the highest velocity and acceleration quantiles while turning left. In contrast, the smallest proportion (0.0008) occurred in the lowest velocity quantile and highest acceleration quantile while moving backward. A detailed toy example in Appendix~\ref{appendix:toy_example} provides a step-by-step explanation of the quantile cube construction process.

The quantile cube can be represented either in deciseconds of time or as proportions of total time spent in each segment. The decisecond representation reflects the absolute time spent in each movement category, whereas the proportional representation captures the athlete's movement distribution across the velocity-acceleration-angle space. The results are organized into an $n \times d$ matrix $\mathbf{Y}$, where each of the $n = 396$ rows corresponds to a vectorized quantile cube from an individual athlete-match-half. Each row contains $d = 100$ features, with each entry indicating the time spent (in deciseconds) within the corresponding movement quantile. The dimensionality ($d = 100$) is defined by the Cartesian product of quantile bins across features: 5 velocity quantiles $\times$ 5 acceleration quantiles $\times$ 4 angle quantiles $= 100$ total combinations. Each feature therefore corresponds to a unique combination of these bins, capturing the joint distribution of movement intensity and direction. The data preprocessing steps from raw GPS data to the quantile cube representation are summarized in the flowchart provided in Appendix~\ref{appendix:flowchart} (Figure~\ref{fig:preprocessing_flowchart}).

\subsection{Covariates}
\label{sec:data_covariates}

In addition to the \gls{gps} data, covariates associated with each match and athlete were obtained, forming an $n \times r$ matrix $\mathbf{X}$ with $n = 396$ rows corresponding to athlete-match-half observations and $r = 13$ covariates. The ten match-level covariates included match ID, the location (home, away, neutral), half ($1^{st}$ or $2^{nd}$), result (win, loss, or tie), goals scored at halftime and full time, goals conceded at halftime and full time, and score differential at halftime and full time. The three athlete-level covariates included the athlete ID, position (defender, midfielder, forward), and playing time by half. 

\section{Methods and Results}
\label{sec:methods_results}

Our analysis of the constructed quantile cubes followed a structured three-step pipeline designed to systematically characterize and model athlete movement patterns. In Section~\ref{sec:halves}, we quantified differences in movement distributions between the first and second halves of each match for every athlete using the Hellinger distance metric. This step captured temporal changes in external load profiles across match halves. In Section~\ref{sec:PCA}, \gls{pca} was applied to the 100-dimensional quantile cube to reduce dimensionality while preserving key variation and to identify dominant movement patterns. Finally, in Section~\ref{sec:DMR}, \gls{dmr} was employed to model the probabilistic relationships between movement distributions and relevant covariates, including player position, playing time, and match factors. The following subsections provide detailed descriptions of the methods and results for each step, and a summary of the notation used is provided in the Notation Table (see Appendix~\ref{appendix:notation}).

\subsection{Quantifying Differences in Movement Distributions Between Match Halves}
\label{sec:halves}

\begin{figure}[ht]
    \centering
    \includegraphics[scale=0.6]{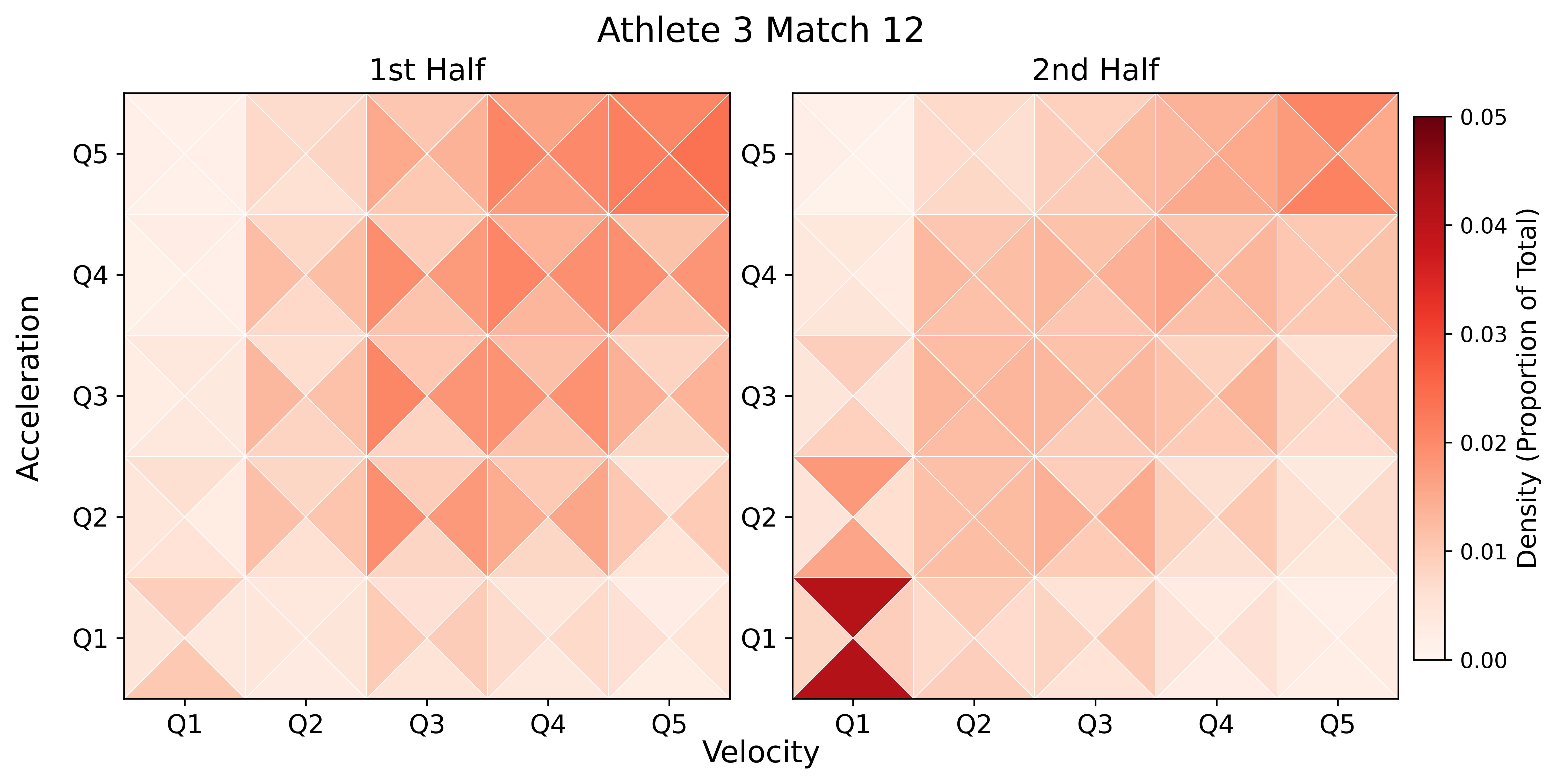}
    \caption{Quantile cubes for Athlete 3 in the first (left) and second (right) halves of Match 12, illustrating shifts in the distribution of velocity and acceleration across the match.}
    \label{fig:distribution_motivation}
\end{figure}

This section quantifies changes in athletes’ movement patterns between the first and second halves of matches by comparing their underlying movement distributions. Due to the high-dimensional and complex nature of the data, classical tests for distributional differences are inappropriate. As illustrated in Figure~\ref{fig:distribution_motivation}, an athlete may spend a higher proportion of time in the higher velocity and acceleration quantiles during the first half, whereas in the second half, the athlete spends more time in the lower velocity and acceleration quantiles. This example highlights shifts across the entire distribution and motivates the use of a distributional metric to capture these changes.

Several existing parametric methodologies are available to measure differences in distributions or means between samples, such as the \textit{t}-test, ANOVA, and Hotelling's $T^2$. However, these tests assume certain conditions, such as common variance, independence, and multivariate normality, which are not satisfied by our data \citep{casella_berger_2002}. Moreover, focusing solely on changes in the mean overlooks the full range of fluctuations contributing, including extreme values that may drive critical changes in external load. High-dimensional statistics addresses scenarios where the number of variables $r$ exceeds the sample size $n$. For example, \citet{bai_saranadasa_1996} introduce a high-dimensional two-sample test that adjusts for the breakdown of classical methods under such conditions, and \citet{chen_quin_2010} developed a test specifically designed for high-dimensional applications such as gene-set testing, where $r$ can be arbitrarily large. Although effective in extreme high-dimensional settings, these approaches still emphasize aggregate changes and may overlook important extreme fluctuations. Therefore, in our moderate-dimensional setting, alternative methodologies are needed to capture the full distribution of movement data, including these extremes.

Given these considerations and the multinomial-nature of the quantile cube data, we propose using the Hellinger distance metric to compare distributions \citep[pp.211-212]{Vaart_1998}. The Hellinger distance is a true metric for measuring the difference between two probability distributions. If $P = (p_1, \ldots, p_d)$ and $Q = (q_1, \ldots, q_d)$ are discrete probability distributions defined on the same finite set $\{1,\ldots,d\}$, then the Hellinger distance is given by
\\
\begin{equation}
    H(P,Q) = \sqrt{ \tfrac{1}{2} \sum_{i=1}^d \left(\sqrt{p_i} - \sqrt{q_i}\right)^2 }.
\end{equation}
\\
Intuitively, the Hellinger distance provides a single number summarizing how different two distributions are. Values close to zero indicate similar halves, while larger values indicate greater differences.

The Hellinger distance metric offers several advantages over alternative metrics, such as the Kullback-Leibler divergence. Its symmetry and boundedness ($0 \leq H(P, Q) \leq 1$) facilitate interpretable and robust comparisons, and its formulation using square roots makes it particularly suitable for multinomial settings \citep[pp.211-212]{Vaart_1998}. The square root transformation naturally moderates the influence of variance across categories, downweighting differences arising from high-variance or low-count bins. This variance-adapting property ensures the stability and meaningfulness of our summaries even when multinomial counts differ, making the Hellinger distance an optimal choice for assessing distributional differences between match halves.

For the analysis, for each athlete $a \in \{1, \dots, 9\}$ and match $m \in \{1, \dots, 23\}$, movement distributions for the first and second halves, denoted by $\hat{p}^{(1)}_{a,m}$ and $\hat{p}^{(2)}_{a,m}$, were estimated. Each $\hat{p}^{(i)}_{a,m}$ is a row of $\mathbf{Y}$, i.e., a $d$-dimensional vector of non-negative values summing to one, representing the proportions of time spent in each cell of the quantile cube. Let $P^{(i)}_{a,m}$ denote the underlying probability distribution of $\hat{p}^{(i)}_{a,m}$, corresponding to the quantile cube for athlete $a$ in match $m$ during half $i \in \{1,2\}$. A formal hypothesis test was applied to determine whether the movement distributions in the first and second halves were statistically equivalent:
\begin{itemize}
    \item \textbf{Null hypothesis ($\mathbf{H_0}$)}: The distributions are the same, i.e., $P_{a,m}^{(1)} = P_{a,m}^{(2)}$.
    \item \textbf{Alternative hypothesis ($\mathbf{H_1}$)}: The distributions differ, i.e., $P_{a,m}^{(1)} \neq P_{a,m}^{(2)}$.
\end{itemize}

\noindent The observed test statistic for each athlete-match pair was
$$\lambda_{a,m} = H(\hat{p}^{(1)}_{a,m}, \hat{p}^{(2)}_{a,m}).$$ Hypothesis testing was conducted using a resampling procedure based on the Hellinger distance. For each pair $(a, m)$, let $t_1$ and $t_2$ denote the athlete's playing time in deciseconds in the first and second halves, respectively. New count vectors of sizes $t_1$ and $t_2$ were simulated by sampling without replacement from the overall movement distribution estimated from all athlete-match-halves, generating simulated first- and second-half samples under the null hypothesis of no distributional difference. For each simulation, the corresponding $d$-dimensional proportion vectors $\hat{p}^{(1)}$ and $\hat{p}^{(2)}$ were calculated from the simulated data, and the Hellinger distance was computed. Repeating this process 10,000 times produced an empirical null distribution of Hellinger distances for each athlete-match pair.

To control the \gls{fwer} due to multiple comparisons, the Bonferroni correction was applied \citep[p.72]{kaltenbach2012concise} with a significance threshold of $\alpha_{a} = 0.05/g_a$, where $g_a$ is the number of valid matches played by athlete $a$. The critical value $c_{a,m}$ for each athlete-match pair was defined as the $(1-$$\alpha_{a})$ quantile of the null distribution. If $\lambda_{a, m} > c_{a, m}$, the null hypothesis was rejected for that athlete-match pair.

\begin{figure}[H]
    \centering
    \includegraphics[scale=0.25]{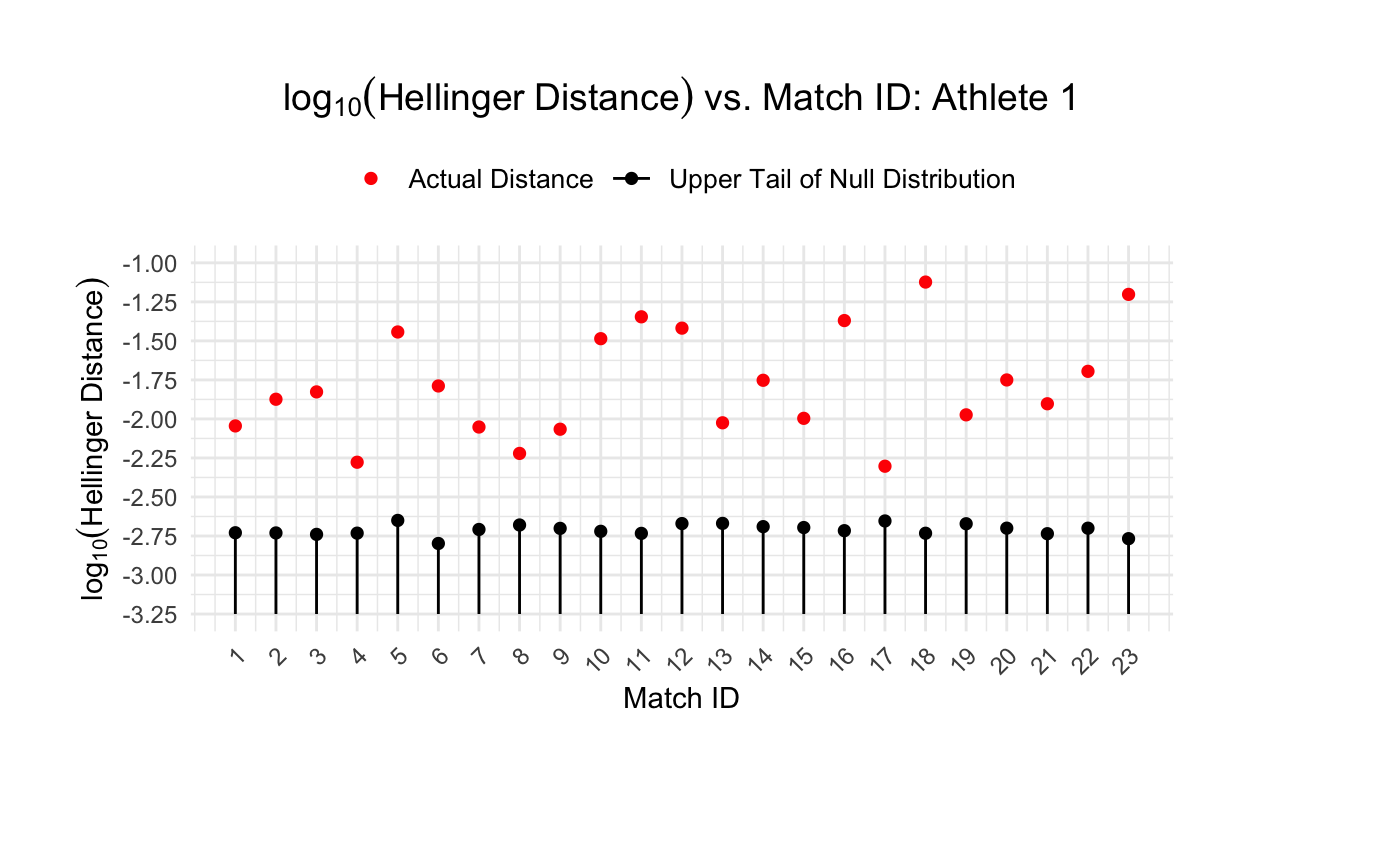}
    \caption{Hellinger distance by match ID between the first and second halves for the actual match data (red circles) and the upper bound of a $\left((1-({0.05}/{23}))\cdot 100 \right)\%$ confidence interval from the null distribution (black circles) for Athlete 1. For all 23 matches, the observed distances exceed the upper bound, indicating that first- and second-half movement distributions differ.}
    \label{fig:athlete5DistributionalDifferences}
\end{figure}

\begin{figure}[H]
    \centering
    \includegraphics[scale=0.5]{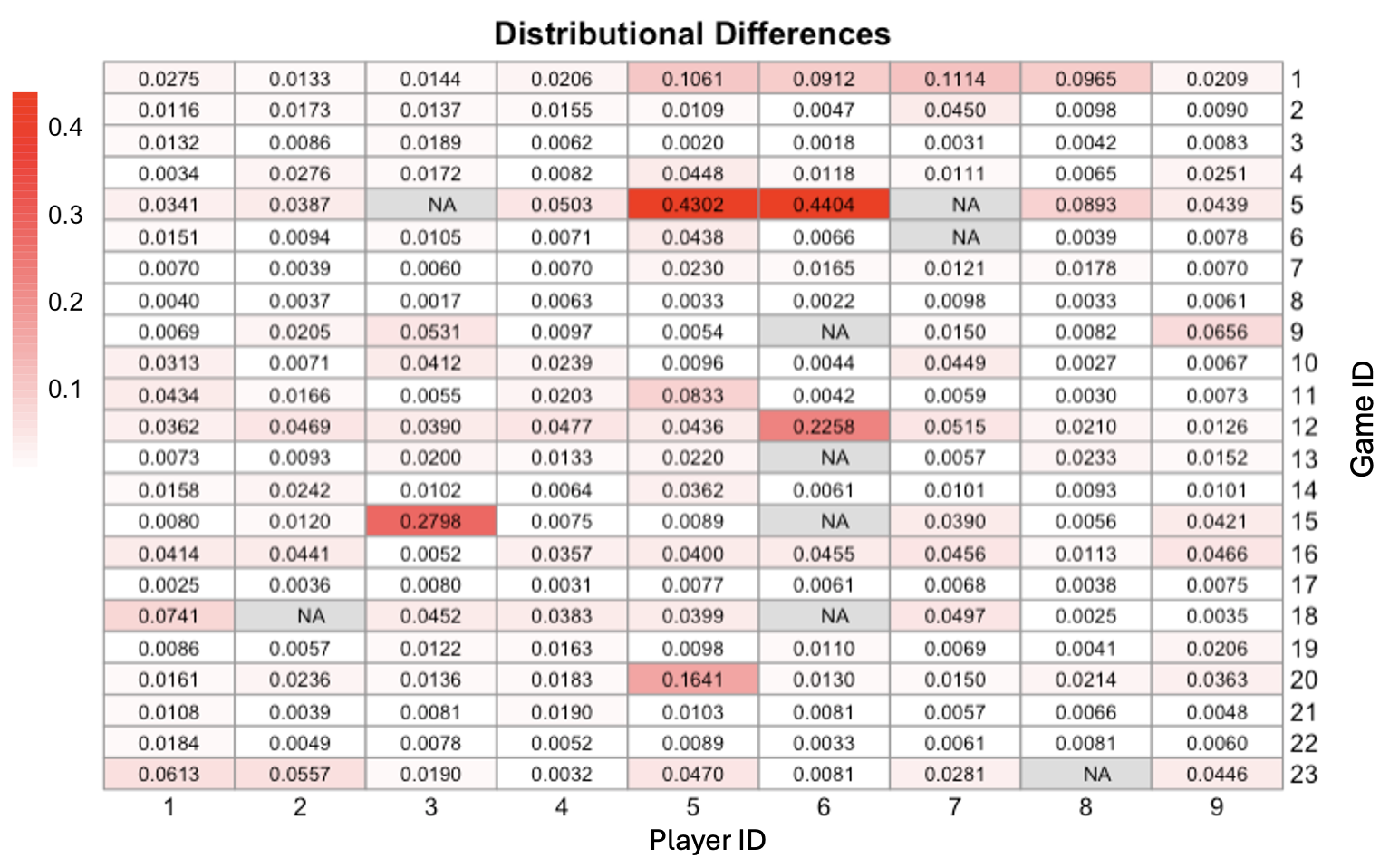}
    \caption{Difference between the Hellinger distance for actual game data and the null distribution at the $(1-\alpha_{a})^{th}$ quantile for all athletes across all matches. NA indicates that the athlete's playing time did not meet the selection criteria for the match ID. The cells are colored according to the difference value, with larger values (darker red) indicating greater differences between the actual match and null distribution. Exact difference values are also provided.}
    \label{fig:distributionalDifferences}
\end{figure}

The hypothesis tests showed significant distributional differences for every athlete-match pair, even after applying the conservative Bonferroni correction. All matches exhibited significant differences between first- and second-half movement distributions, with no cases where halves were statistically indistinguishable after multiple comparison adjustment. Figure~\ref{fig:athlete5DistributionalDifferences} illustrates these results for Athlete 1, and Figure~\ref{fig:distributionalDifferences} summarizes the findings across all athletes using the Bonferroni correction.

\subsection{Dimensionality Reduction of Movement Patterns in Match Contexts}
\label{sec:PCA}

To identify the dominant patterns underlying athlete movement behaviors and to reduce the computational complexity of the 100-dimensional quantile cube data, \gls{pca} was applied to the observed count matrix $\mathbf{Y}$ (defined in Section~\ref{sec:data_quantile_cube}). This dimensionality reduction approach allows extraction of the key modes of variation, summarizing movement distributions while providing interpretable insights into match-specific and athlete-specific dynamics \citep{jolliffe2002principal}. The resulting \gls{pcs} offer a structured framework for detecting anomalous movement patterns, distinctive match characteristics, and systematic variations in external load profiles across different competitive contexts.

\begin{figure}[H]
    \centering
    \includegraphics[scale=0.25]{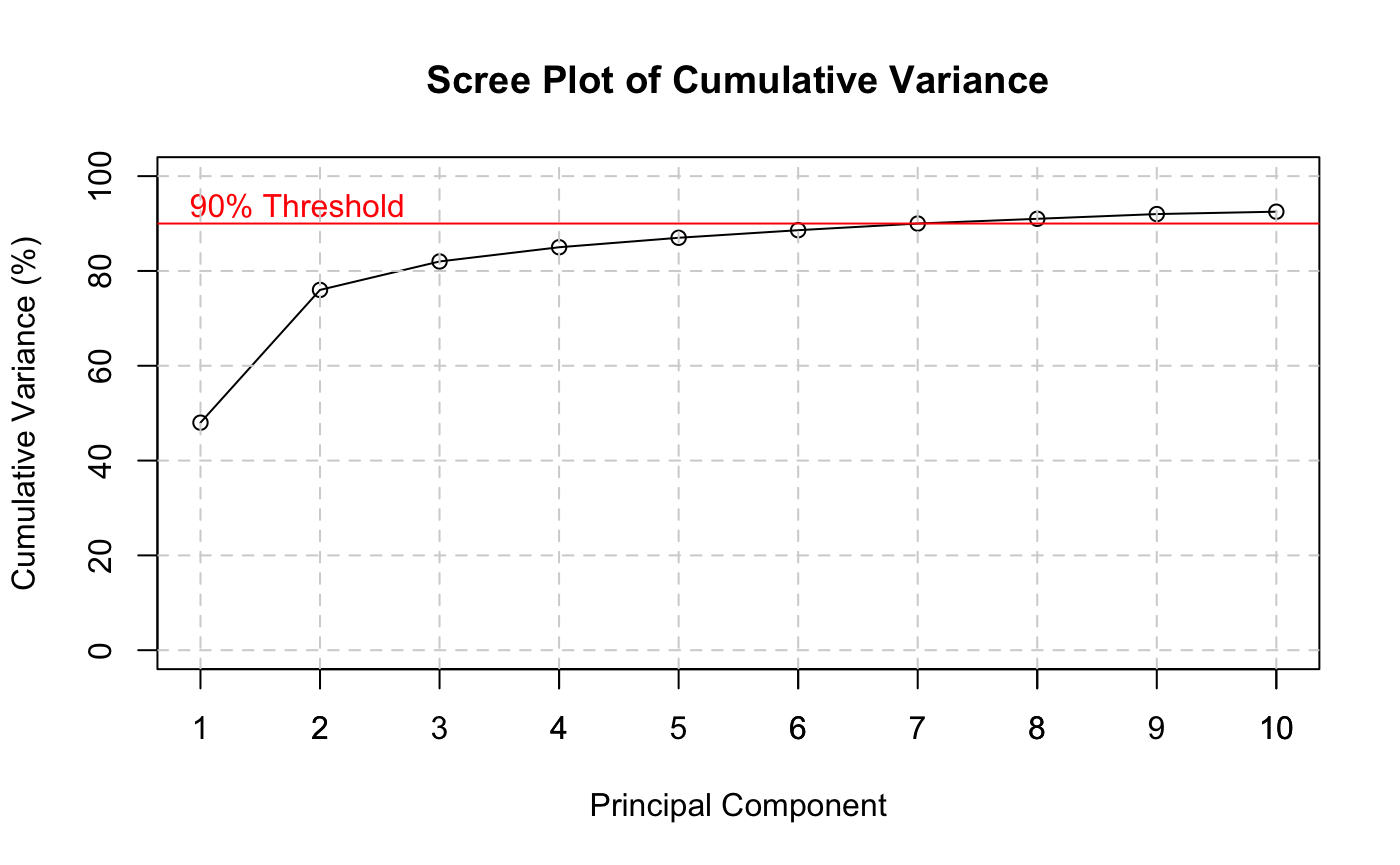}
    \caption{Cumulative variance explained by the top 10 \gls{pcs} from the quantile cube analysis. The decline in explained variance after the 7th component supports the selection of a lower-dimensional representation for downstream analysis.}
    \label{fig:screeplot}
\end{figure}

\gls{pca} decomposition of $\mathbf{Y}$ yielded 100 \gls{pcs}, each linked to a 396-dimensional score vector representing the projection of individual athlete-match-halves onto the component space. To determine the optimal number of components for downstream analysis, a cutoff of 90\% variance explained was applied, resulting in the retention of the first seven \gls{pcs} (Figure~\ref{fig:screeplot}). This criterion ensures that the reduced representation captures the majority of systematic variation while minimizing noise and redundant information from lower-variance components.

However, variance explained alone does not guarantee the practical interpretability or actionable relevance of the \gls{pcs} for understanding athlete performance. To assess their meaningfulness, component scores were plotted against match- and athlete-level characteristics from the design matrix $\mathbf{X}$ (defined in Section 2.2). While \gls{pc2} and \gls{pc3} explained substantial variance, their loadings were diffusely distributed across velocity-acceleration-angle bins, showing no coherent patterns linked to specific covariates. Scatterplots confirmed no discernible clustering or separation, indicating these components primarily capture subtle, distributed variations rather than systematic behavioral differences. In contrast, \gls{pc1} and \gls{pc4} demonstrated clear interpretability, exhibiting clustering patterns directly associated with specific matches and movement dynamics. Therefore, only \gls{pc1} and \gls{pc4} are presented here as they provided the most actionable insights into external load variations. 

\begin{figure}[H]
    \centering
    \includegraphics[scale=0.6]{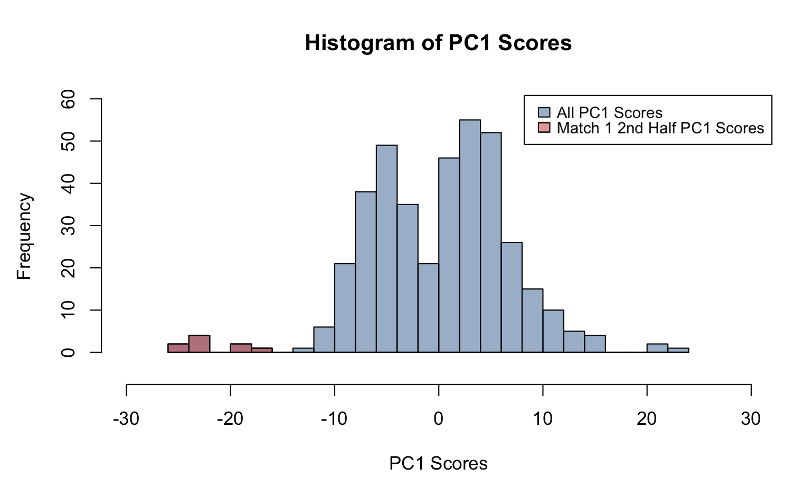}
    \caption{Histogram comparing the distribution of \gls{pc1} scores for all athlete-match-halves (blue) versus the second half of Match 1 (red). Scores from the second half of Match 1 are shifted toward more negative values relative to the overall distribution, indicating reduced time spent in the quantiles most strongly associated with this component during that period.}
    \label{fig:histogramPC1}
\end{figure}

\gls{pc1}, accounting for the largest proportion of variance, captured distinctive movement characteristics observed during the second half of Match 1, with loadings revealing a systematic reduction in time spent in the middle quantiles of velocity and acceleration. This indicates a shift toward more polarized movement patterns, characterized by either low-intensity positioning or high-intensity bursts, with less time spent in moderate-intensity activities. Table~\ref{table:pca_loadings_pc1} presents the ten highest-magnitude loadings for \gls{pc1}, dominated by features combining the third velocity quantile (moderate running speeds) with the fourth acceleration quantile (high acceleration) across forward and backward movement directions (first and third angle quantiles). This negative loading pattern reflects reduced time spent in movement categories requiring moderate velocity paired with high acceleration. The uniqueness of this pattern is further confirmed by comparing \gls{pc1} scores across all athlete-match-halves: Figure~\ref{fig:histogramPC1} shows that observations from Match 1's second half were systematically more negative relative to the overall distribution, highlighting a significant deviation from typical movement patterns observed throughout the season.

\begin{figure}[H]
    \centering
    \includegraphics[scale=0.6]{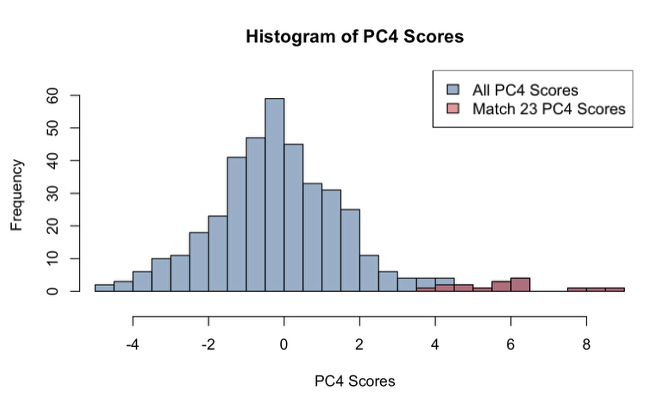} 
    \caption{Histogram comparing the distribution of \gls{pc4} scores for all athlete-match-halves (blue) versus Match 23 (red). The Match 23 \gls{pc4} scores are shifted toward higher values relative to the overall distribution, suggesting this match involved increased time spent in low-velocity, high-acceleration quantiles associated with this component.}
    \label{fig:histogramPC4}
\end{figure}

\gls{pc4} captured the distinctive movement characteristics of Match 23, the final match of the season. Table~\ref{table:pca_loadings_pc4} presents the component's highest-magnitude loadings, characterized by combinations of low velocity (first and second quantiles) with maximal acceleration (fifth quantile), particularly in the forward and backward directions. These positive loadings indicate increased time spent in low-velocity, high-acceleration movements throughout the whole match. Figure~\ref{fig:histogramPC4} confirms that \gls{pc4} scores for Match 23 were systematically elevated relative to the season-long distribution, highlighting the uniqueness of the final match’s movement patterns.

\begin{table}[ht]
\centering
\begin{minipage}{0.45\textwidth}
\footnotesize
\begin{tabular}{|c|c|c|}
\hline
\textbf{} & \textbf{Variable} & \textbf{PC1 Loading} \\
\hline
1 & Q3\_vel\_Q4\_acc\_Q3\_angle & -0.1306 \\
2 & Q3\_vel\_Q4\_acc\_Q1\_angle & -0.1300 \\
3 & Q4\_vel\_Q3\_acc\_Q3\_angle & -0.1293 \\
4 & Q4\_vel\_Q3\_acc\_Q1\_angle & -0.1290 \\
5 & Q4\_vel\_Q4\_acc\_Q1\_angle & -0.1273 \\
6 & Q4\_vel\_Q4\_acc\_Q3\_angle & -0.1271 \\
7 & Q4\_vel\_Q2\_acc\_Q4\_angle & -0.1253 \\
8 & Q4\_vel\_Q2\_acc\_Q1\_angle & -0.1253 \\
9 & Q2\_vel\_Q5\_acc\_Q2\_angle & -0.1253 \\
10 & Q3\_vel\_Q3\_acc\_Q3\_angle & -0.1253 \\
\hline
\end{tabular}
\caption{Top 10 absolute loadings for \gls{pc1}}
\label{table:pca_loadings_pc1}
\end{minipage}%
\hspace{0.5cm} 
\begin{minipage}{0.45\textwidth}
\footnotesize
\begin{tabular}{|c|c|c|}
\hline
\textbf{} & \textbf{Variable} & \textbf{PC4 Loading} \\
\hline
1 & Q1\_vel\_Q5\_acc\_Q1\_angle & 0.3297 \\
2 & Q1\_vel\_Q5\_acc\_Q3\_angle & 0.3179 \\
3 & Q2\_vel\_Q5\_acc\_Q1\_angle & 0.2215 \\
4 & Q2\_vel\_Q5\_acc\_Q3\_angle & 0.2111 \\
5 & Q1\_vel\_Q5\_acc\_Q4\_angle & 0.2053 \\
6 & Q1\_vel\_Q5\_acc\_Q2\_angle & 0.2022 \\
7 & Q4\_vel\_Q5\_acc\_Q3\_angle & 0.1734 \\
8 & Q3\_vel\_Q5\_acc\_Q3\_angle & 0.1709 \\
9 & Q3\_vel\_Q5\_acc\_Q1\_angle & 0.1692 \\
10 & Q4\_vel\_Q5\_acc\_Q1\_angle & 0.1642 \\
\hline
\end{tabular}
\caption{Top 10 absolute loadings for \gls{pc4}}
\label{table:pca_loadings_pc4}
\end{minipage}
\end{table}

\gls{pca} successfully reduced the dimensionality from 100 to 7 components, retaining 90\% of the variance in the quantile cube data. Among these, \gls{pc1} and \gls{pc4} showed the strongest associations with specific matches and interpretable movement patterns, with \gls{pc1} reflecting the distinctive characteristics of Match 1 and \gls{pc4} reflecting those of Match 23.

\subsection{Modeling Movement Distributions as a Function of Player and Match \\Characteristics}
\label{sec:DMR}

Having established systematic distributional differences between match halves and identified key movement patterns through dimensionality reduction, the next step was to quantify how these movement distributions varied as a function of player and match characteristics using formal statistical modeling. To characterize the relationships between athlete movement distributions and contextual factors, \gls{dmr}, a flexible modeling framework specifically designed for compositional count data with overdispersion, was employed. The quantile cube data exhibit two key features that necessitate this specialized approach: (1) compositional dependence, where time allocated to one movement category directly constrains time available for others; and (2) overdispersion, where the observed variance in movement category counts exceeds the variance predicted by standard multinomial models due to individual athlete differences, match-specific contexts, and temporal clustering effects.

The \gls{dm} distribution addresses both features by modeling the underlying category probabilities as random draws from a Dirichlet distribution, naturally accommodating the compositional constraints while allowing greater variance than multinomial models \citep{mosimann_compound_1962, chen_variable_2013}. The \gls{dmr} framework extends this model to a regression setting, enabling systematic incorporation of external covariates such as match half, player position, and playing time. This approach provides a principled method to quantify how movement patterns vary with contextual factors while respecting the inherent structure of compositional movement data.

\subsubsection{Model Specification}
\label{sec:DMR_model_specification}

For each observation (row) $i$ in the count matrix $\mathbf{Y}$, the movement distribution $\mathbf{y_i} = (y_{i1}, y_{i2},\dots, y_{id})$ was modeled using the \gls{dm} distribution:
\\
\begin{equation}
    f_{DM}(\mathbf{y_i}|\boldsymbol{\eta}) = \frac{\Gamma(N_i + 
1)\Gamma(\sum_{j=1}^d \eta_j)}{\Gamma(N_i + \sum_{j=1}^d \eta_j)\prod_{j=1}^d \frac{\Gamma(y_{ij} + \eta_j)}{\Gamma( \eta_j)\Gamma(y_{ij} + 1)}}.
\end{equation}
\\ 
where $\boldsymbol{\eta} = (\eta_1, \eta_2, \dots, \eta_d)$ are positive concentration parameters for each movement category and $N_i = \sum_{j=1}^d y_{ij}$ is the total movement time for athlete-match-half $i$
\citep{mosimann_compound_1962, chen_variable_2013}. These $\eta_j$ are unknown parameters of the model to be estimated from the observed count data. Intuitively, they control the expected proportions and variability of each movement category.

To incorporate covariate effects, each concentration parameter $\eta_j$ for movement category $j \in \{1, \dots, d\}$ was modeled using a log-linear formulation:
\\
\begin{equation} \label{alphaLogLinearModel}
    \eta_j := \exp\left(\beta_{j0} + \sum_{k=1}^r \beta_{jk}x_{ik}\right),
\end{equation}
\\
where $\beta_{j0}$ represents the baseline for movement category $j$, $\beta_{jk}$ quantifies the effect of covariate $k$ on category $j$, and $x_{ik}$ denotes the value of covariate $k$ for observation $i$ \citep{mosimann_compound_1962, chen_variable_2013}. This formulation makes the role of $\eta_j$ explicit, modeling them as functions of covariates rather than fixed values. This ensures positivity while allowing flexible, interpretable relationships between covariates and movement patterns across all $d = 100$ quantile cube categories. Put simply, this regression framework allows us to see how different factors, such player position, match result, or playing time, affect the overall distribution of an athlete’s movements, while accounting for the fact that time spent in one type of movement limits time available for others and that real data are more variable than a simple model would assume.

\subsubsection{Covariate Encoding and Selection}
\label{sec:DMR_covariates}

All covariates in the design matrix $\mathbf{X}$ were included as potential predictors in the regression model. Categorical variables were encoded using standard dummy-variable approaches: binary factors (e.g., match half) as 0/1 indicators, and multi-level factors (e.g., athlete position, match location) with one reference level omitted. Playing time was mean-centered and log-transformed to improve interpretability and stabilize estimation.

All candidate models were estimated using the \texttt{MGLMfit} function from the MGLM package in R \citep{MGLM_journal, MGLM_R}, which applies maximum likelihood estimation to obtain the concentration parameters $\eta_j$ through the log-linear regression coefficients $\beta_{jk}$ in Equation~$\ref{alphaLogLinearModel}$. Candidate models were systematically compared, testing different combinations of covariates. The optimal model, selected for both statistical performance and interpretability, included three key predictors: match half ($1^{st}, 2^{nd}$), player position (defender, midfielder, forward), and mean-centered $\log(\text{playing time})$. This specification allowed the detection of systematic changes in movement patterns across these primary contextual factors while maintaining model simplicity.

\subsubsection{Model Results}
\label{sec:DMR_model_results}

Parameter estimation identified significant covariate effects across multiple movement categories. Coefficients from Equation~\ref{alphaLogLinearModel} were considered statistically significant when \\\mbox{$\big|\beta_{jk}/SE_{\beta_{jk}}\big| > 3$}, where $SE_{\beta_{jk}}$ is the standard error of $\beta_{jk}$, computed from the observed Fisher information matrix provided by \texttt{MGLMfit} \citep{MGLM_journal, MGLM_R}. This threshold was determined based on the empirical distribution of standardized coefficients in the dataset, offering a conservative approach to highlight meaningful effects.

\begin{figure}[H]
\centering
    \includegraphics[width=0.45\textwidth]{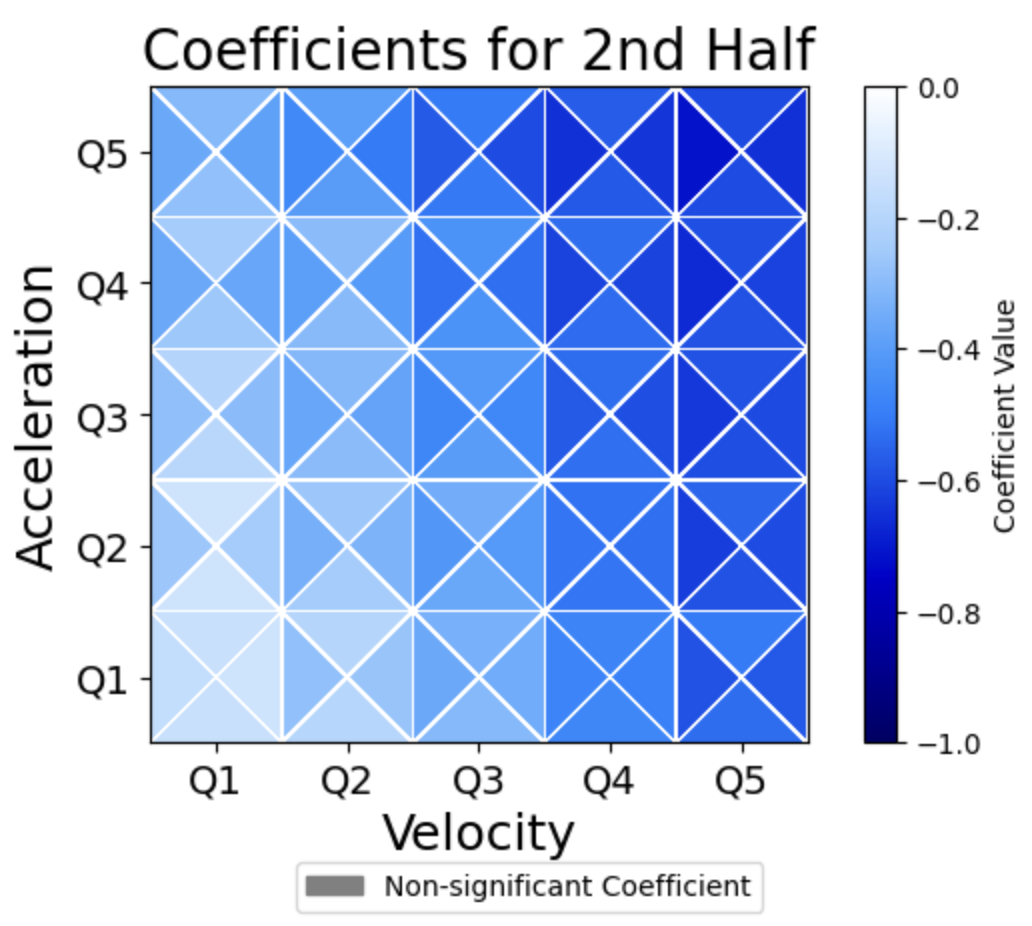}
    \includegraphics[width=0.45\textwidth]{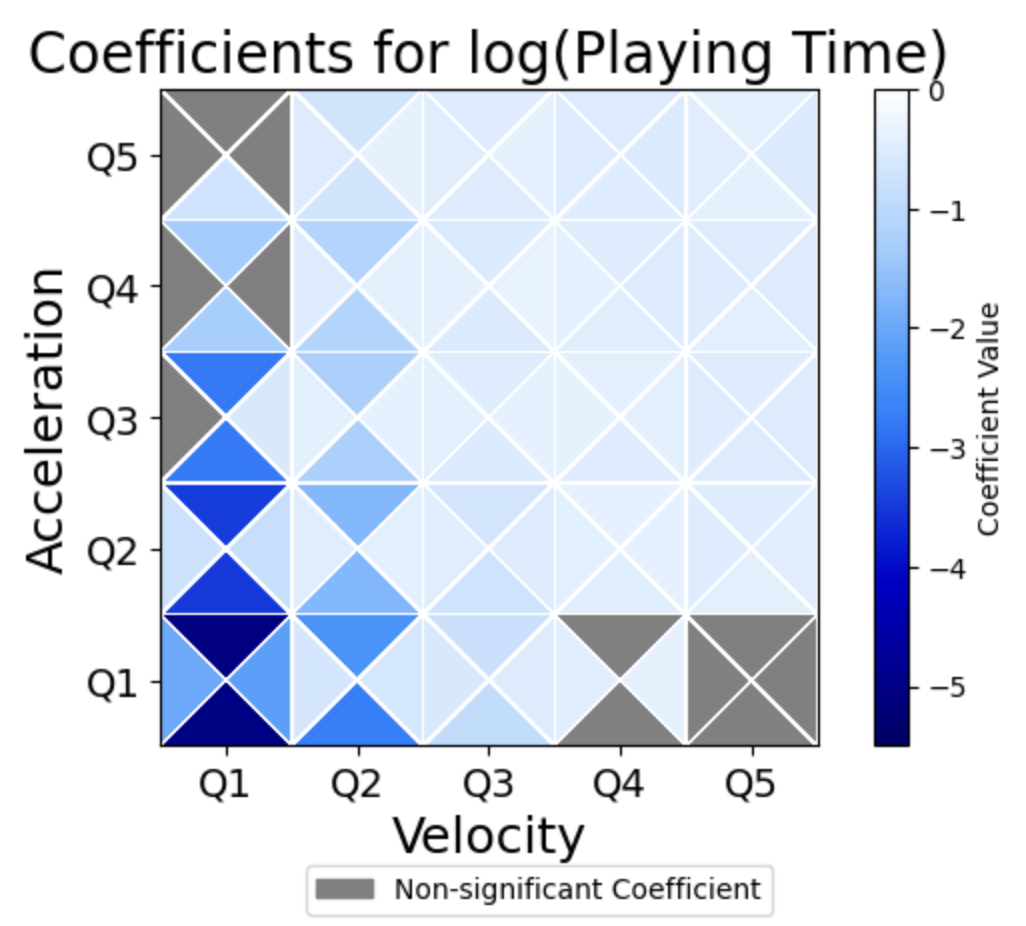}
    \caption{Quantile cube illustrating the \gls{dmr}} coefficients for the second half (left) and log(Playing Time) (right). Non-significant coefficients are shown in gray, while significant coefficients are color-coded based on effect size, with intensity indicating magnitude.
    \label{fig:coefsHalfTime}
\end{figure}

Match half and playing time effects (Figure~\ref{fig:coefsHalfTime}) revealed systematic changes in movement patterns. During the second half, more negative coefficients dominated the higher velocity quantiles across acceleration and angle categories, indicating decreased time spent in high-intensity movement patterns as matches progressed, consistent across all movement directions. Additionally, athletes with above-average playing time exhibited distinct movement signatures compared to those with shorter durations, spending significantly less time in the lowest velocity and acceleration quantiles, particularly in forward and backward directions, suggesting that longer-playing athletes maintain higher baseline activity levels throughout the match.

\begin{figure}[H]
    \centering
    \includegraphics[width=0.45\textwidth]{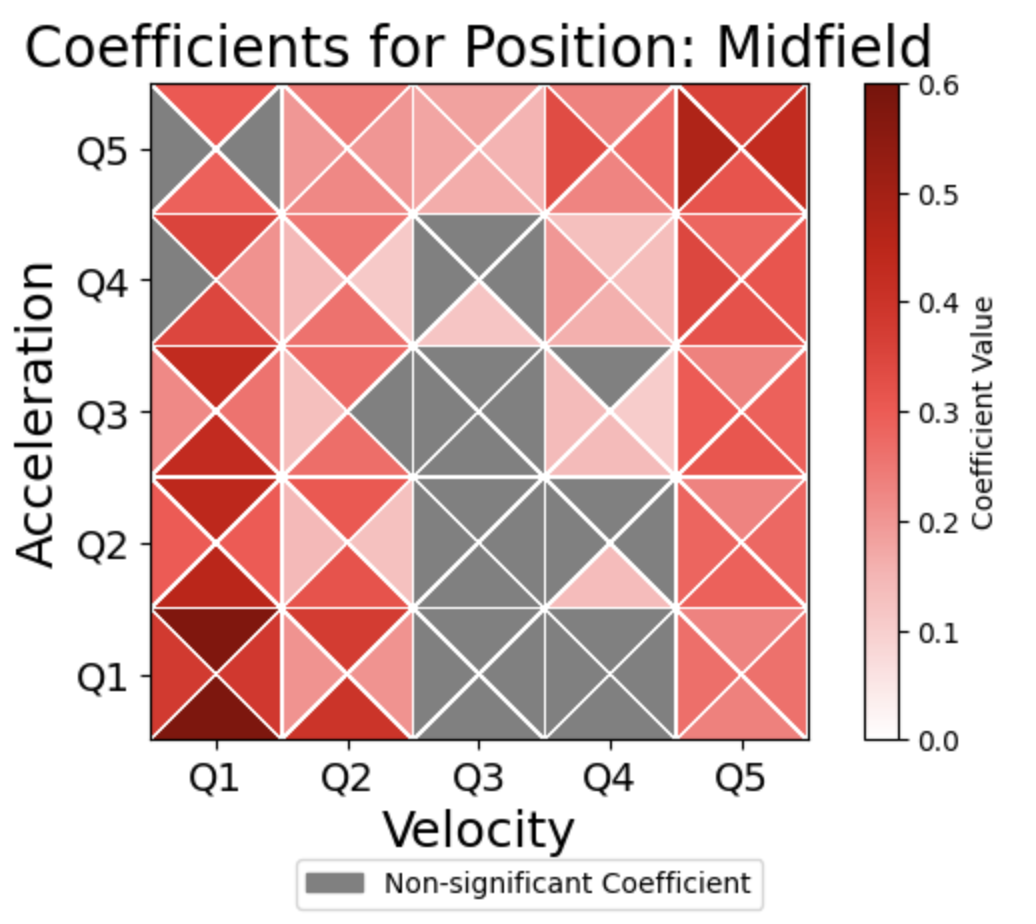}
    \includegraphics[width=0.45\textwidth]{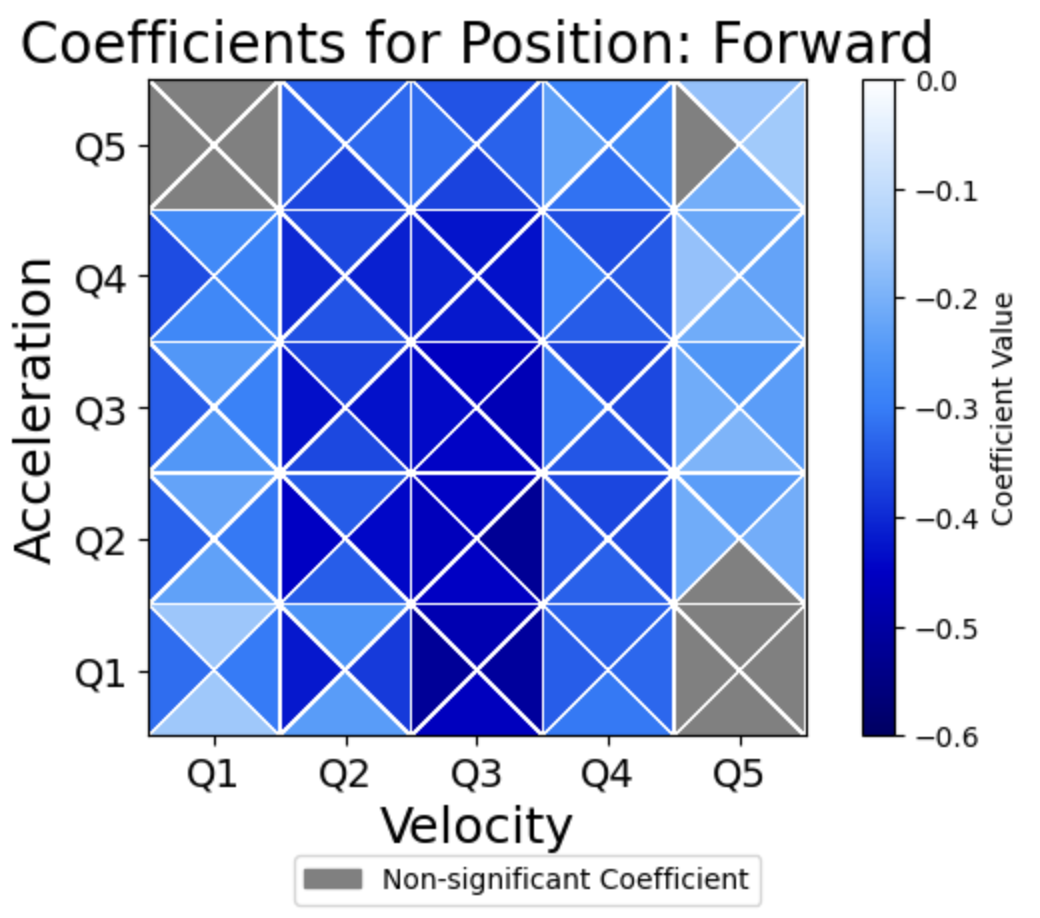}
    \caption{Quantile cube depicting the \gls{dmr} coefficients for player position, where the defender is treated as the base class. Coefficients for comparisons with midfielders are shown on the left, and those for forwards are shown on the right. Non-significant coefficients are shaded in gray, while significant coefficients are color-coded based on effect size and direction: blue for negative effects and red for positive effects, with intensity reflecting magnitude.}
    \label{fig:coefsPosition}
\end{figure}

Positional differences exhibited clear and interpretable patterns (Figure~\ref{fig:coefsPosition}). Compared to defenders (reference category), midfielders spent more time in lower velocity and acceleration quantiles, combined with elevated time in the highest velocity quantile across all acceleration levels. This bimodal pattern suggests midfielders alternate between periods of lower-intensity positioning and high-intensity running. Forwards demonstrated a contrasting pattern, spending less time in middle and lower velocity quantiles across most acceleration categories, with one notable exception in the first velocity quantile at maximum acceleration. Additionally, forwards exhibited significantly reduced left-right movement compared to forward-backward movement relative to defenders.
 
\section{Discussion}
\label{sec:discussion}

This paper presents a novel adaptation of established methodologies for assessing external load in elite female soccer athletes, with a focus on improving the interpretability of movement patterns during match play. By leveraging wearable \gls{gps} data, we examined the relationships between velocity, acceleration, and angle of movement with athlete and match characteristics. Our findings demonstrate how a probabilistic, distribution-based analytic framework can uncover patterns that conventional metrics overlook, providing new insights into performance and fatigue in elite women’s soccer.

Our analysis revealed significant differences in athlete movement patterns between the first and second halves of matches. Specifically, within our sample of elite women’s soccer athletes, quantile cube distributions differed significantly between halves, reinforcing the influence of match duration on movement dynamics. This finding aligns with previous research, such as \citet{barrera_2021}, which reported reductions in external load metrics like high-speed running during the second half of professional male soccer matches, and uses a distributional lens to extend \citeauthor{barrera_2021}'s conclusions to women’s soccer. These results suggest that, in this dataset, second-half differences were not confined to one or two performance metrics but reflected a broader reshaping of movement intensity profiles that may be driven by neuromuscular fatigue or tactical decisions.

Additionally, the use of PCA to reduce the dimensionality of the quantile cubes revealed context-specific deviations across the season. Most notably, the second half of Match 1 (season opener) and Match 23 (postseason tournament match) displayed movement profiles distinct from typical seasonal patterns. Match 1 did not involve a major rival and may reflect early-season conditioning rather than tactical pressure, whereas Match 23, a decisive postseason loss, probably reflects accumulated fatigue combined with heightened match intensity. These deviations point to contextual factors such as early-season conditioning, late-season fatigue, or tactical adaptations. This highlights the importance of considering both temporal (within-match) and contextual (seasonal, environmental, or competitive) factors in workload analysis. For instance, environmental conditions like moderate altitude reduce high-intensity running and overall distance in collegiate female soccer matches \citep{Bohner2015}, while workload demands vary between in-conference and out-of-conference play \citep{Bozzini2020}. Such findings underscore the value of incorporating context directly within analytic frameworks rather than relying on fixed reference values.

The \gls{dmr} model confirmed substantial differences between first and second halves, with athletes spending less time in the higher quantiles of both velocity and acceleration as the match progressed, providing statistical evidence of second-half intensity decline. This pattern likely reflects acute fatigue or strategic pacing \citep{Snyder2024, Andersson2008, barrera_2021}. Importantly, athletes with greater playing time per half spent less time in the lower velocity and acceleration bins, suggesting these players maintain a higher baseline workload despite extended minutes, an encouraging indicator for endurance and load management strategies.

The coefficients of the \gls{dmr} also revealed positional differences for midfielders and forwards in relation to defenders. Midfielders exhibited a bimodal load profile, alternating between low-intensity positioning and high-intensity bursts. This dynamic role is consistent with prior work in women’s and men’s soccer showing that midfielders cover greater total distance and wider intensity ranges than defenders \citep{Vescovi2014, wehbe_2014, Panduro2022}. Forwards, on the other hand, spent significantly less time in middle velocity quantiles across accelerations, particularly with lateral movements, supporting their role in alternating between recovery and forward-directed sprints. Prior research in men’s soccer has shown that defenders perform fewer high-speed runs but sustain notable acceleration demands \citep{wehbe_2014}, highlighting the importance of velocity–acceleration metrics and the limitations of relying solely on distance- or sprint-based measures.

When interpreting our findings, it is essential to consider contextual and competition level factors demonstrated in related research. Seasonal and postseason workload variations indicate changes in running intensity and volume that can impact athlete performance and fatigue \citep{Wells2015}. Competition at international levels presents higher demands in high-speed running and sprints, particularly affecting midfielders and defenders, further supporting the necessity of individualized training and monitoring strategies \citep{Mara2017, Griffin2021, Datson2017}. These considerations, alongside environmental and individual recovery factors, emphasize the complexity of athlete monitoring while highlighting opportunities for integrated, multivariate analytics.

\subsection{Practical Implications}
\label{sec:discussion_practical_implications}

Taken together, these results have important practical implications for athlete monitoring and management. Our work introduces an accessible statistical method that transforms raw \gls{gps} measurements into probabilistic insights about athlete movement patterns. The \gls{dmr} model allows integration of positional and match factors, as well as environmental or physiological covariates, enabling predictions of movement responses under varying conditions. This facilitates evidence-based decisions about training load, recovery, and substitution.

Importantly, we believe these insights can translate directly into future practical applications for coaching and athlete management:
\begin{itemize}
    \item \textbf{Enhanced substitution strategies:} Coaches can use probabilistic movement profiles generated by the model to identify players whose movement patterns deviate significantly from their baseline in real time, signaling acute fatigue or injury risk before overt signs appear.
    \item \textbf{Tailored training prescriptions:} Training drills can be informed by quantified positional movement demands. For example, midfielders exhibiting wide velocity and acceleration distributions may benefit from conditioning emphasizing both endurance and explosive speed, while defenders might focus on drills stressing acceleration bursts and recovery.
    \item \textbf{Context-aware workload management:} Understanding how environmental and match- specific factors affect workload allows practitioners to adjust training intensity and recovery strategies accordingly. For instance, if altitude reduces high-intensity running capacity as suggested by prior research, training loads can be moderated before competitions at such venues \citep{Bohner2015}.
    \item \textbf{Integrated internal and external load monitoring:} By combining movement distributions with athlete wellness data (e.g., soreness, sleep quality), sports science teams can implement individualized recovery protocols and readiness assessments, ultimately promoting injury prevention and sustainable performance.
\end{itemize}

Building on these applied scenarios, implementing this methodology in real time involves integrating continuous \gls{gps} data streams from training and matches, establishing player-specific baselines, and detecting statistically significant deviations from typical movement profiles. This approach elevates \gls{gps} monitoring from simple tracking to a rigorous statistical tool, where decisions about player management are guided by formal inference rather than arbitrary thresholds. In practice, it enables sports science teams to deploy dashboards or automated alert systems that notify coaching staff of atypical movement patterns warranting intervention, supporting proactive strategies such as substitution, load adjustment, or medical evaluation to better safeguard athlete health and performance.

\section{Conclusion}
\label{sec:conclusion}

This study provides a novel, probabilistic framework that advances the analysis of external load in women’s soccer. By introducing the quantile cube approach combined with Hellinger distance, \gls{pca}, and \gls{dmr}, we demonstrate how complex GPS trajectories can be summarized into interpretable, distribution-based measures of movement. These methods capture nuanced variation and enable individualized inference, offering richer insights than traditional aggregate or threshold-based metrics. 

Importantly, our female-specific dataset offers a unique perspective that addresses a critical gap in existing research, which predominantly focuses on male athletes. This study responds to longstanding calls for gender-specific workload analytics and contextual interpretations aimed at enhancing training strategies and reducing injury risk \citep{Mujika2009}. Disparities in physical fitness and performance capacities between genders and competitive levels are well documented, further underscoring the necessity of female-specific data and analytic frameworks like the ones developed here to support effective and tailored workload management.

Despite these strengths, several limitations should be noted. First, our sample size was modest (nine athletes from a single team over one season), limiting the generalizability of findings. The exclusion of wide-vs-central positional subdivisions and goalkeepers further restricts the applicability of findings. Taking into account these positions as separate categories may reveal additional differences in movement patterns, as studies have found that in male soccer athletes, wide position players typically produce higher acceleration efforts than central position players \citep{ingebrigtsen_2015}. Additionally, match session selection was based on a threshold of at least 25 minutes per half; while this ensures data quality, it may introduce selection bias by excluding shorter substitution stints and atypical playing patterns.

Our analysis did not include training sessions, recovery protocols, or internal load factors such as heart rate, perceived exertion, or biochemical markers—key components of comprehensive athlete monitoring. Further, important contextual influences such as weather conditions, opposition strength, fixture congestion, match importance (postseason vs. regular season), and pitch quality were not included in our models, despite evidence that these factors influence external workload. Future research should seek to incorporate altitude and match-type effects \citep{Bohner2015, Bozzini2020}, as these have been shown to affect high-intensity work and overall match demands.

Technological limitations must also be considered: while \gls{gps} devices provide accurate tracking of most movement patterns, they may underperform when capturing abrupt or highly multidirectional changes. The addition of \gls{imus}---which integrate accelerometers and gyroscopes---could enable high-frequency, real-time data collection for more precise analysis of rapid, multidimensional movements that traditional GPS devices may overlook \citep{Mudeng2022}. Advances in monitoring technology now allow measurement of physiological variables (such as heart rate, heart rate variability, and neuromuscular fatigue markers), which would further enrich future workload analyses.

Finally, the selection and binning of quantiles in the quantile cube framework, while aiming for interpretability and robustness, are somewhat arbitrary and may need tuning for other teams or application scenarios. Future research should employ larger, multisite cohorts, include training and recovery data, integrate contextual and physiological covariates, and validate these models longitudinally—including prospective injury and recovery outcomes—to maximize translational relevance for athlete health and performance.

In conclusion, this study demonstrates that a multidimensional, quantile-based approach enhances interpretability, statistical rigor, and practical utility of GPS-derived external load in women’s soccer. By bridging wearable technology, probabilistic modeling, and applied sports science, we establish a methodological foundation for individualized, data-driven athlete management aimed at optimizing performance and reducing injury risk. Continued research should further validate and refine these approaches across larger and more diverse cohorts, linking external load patterns with internal physiology, recovery, contextual variables, and long-term outcomes, to maximize their translational impact on athlete health and performance.

\section*{Acknowledgements}
We extend our gratitude to Elena Cantu, Sam Moore, and Dr. Abbie Smith-Ryan from the Applied Physiology Lab in the University of North Carolina at Chapel Hill’s Department of Exercise and Sport Science for their generous contributions in gathering and providing access
to the data.

\section*{Funding}
Jan Hannig’s research was supported in part by the National Science Foundation under Grant No. DMS-1916115, 2113404, and 2210337.

\bibliography{references}

\appendix

\newpage

\section{Notation Table}
\label{appendix:notation}

\begin{table}[H]
\centering
\begin{tabularx}{\textwidth}{@{} l X @{}}
\toprule
\textbf{Symbol} & \textbf{Description} \\
\midrule
$n$ & Number of observations (athlete-match-halves), $n = 396$ \\
$d$ & Dimension of the quantile cube vector, $d = 5 \times 5 \times 4 = 100$ \\
$r$ & Number of covariates in the regression model, $r = 13$ \\
$i$ & Index for observation $i = 1, \dots, n$ \\
$a$, $m$, $h$ & Athlete ID ($a \in \{1,\dots,9\}$), Match ID ($m \in \{1,\dots,23\}$), and Half ($h \in \{1,2\}$) \\
$t_i$ & Playing time in deciseconds for observation $i$ \\
$\mathbf{X} = (x_{ik})_{n \times p}$ & Design matrix of covariates \\
$x_{ik}$ & Value of covariate $k$ for observation $i$ \\
$k$ & Index for covariate, $k = 1, \dots, 13$\\
$\mathbf{Y} = (y_{ij})_{n \times d}$ & Observed count matrix of quantile cube vectors \\
$y_{ij}$ & Time (in deciseconds) spent in quantile bin $j$ by observation $i$ \\
$j$ & Index for quantile cube bins, $j = 1, \dots, 100$ \\
$H(P, Q)$ & Hellinger distance between two discrete distributions $P$ and $Q$ \\
$\hat{p}^{(1)}_{a,m}$, $\hat{p}^{(2)}_{a,m}$ & Empirical distributions from the quantile cube for first and second halves \\
$\lambda_{a,m}$ & Observed Hellinger distance between halves for athlete $a$ in match $m$ \\
$c_{a,m}$ & Bonferroni-corrected critical value (threshold) from the empirical null distribution for athlete $a$ in match $m$ used in the Hellinger distance test \\
$\pi_j$ & Probability of occupying quantile bin $j$ in \gls{dmr} model \\
$\boldsymbol{\pi} = (\pi_1, \dots, \pi_d)$ & Vector of movement probabilities over quantile bins \\
$\eta_j$ & Dirichlet concentration parameter for quantile bin $j$ \\
$\boldsymbol{\eta} = (\eta_1, \dots, \eta_d)$ & Vector of Dirichlet concentration parameters \\
$\beta_{j0}$ & Intercept for bin $j$ in the DMR model \\
$\beta_{jk}$ & Coefficient for covariate $k$ on bin $j$ \\
\bottomrule
\end{tabularx}
\caption{Summary of notation used throughout the paper.}
\label{tab:notation}
\end{table}

\newpage

\section{Toy Example of a Quantile Cube}
\label{appendix:toy_example}

We present a toy example of a quantile cube to demonstrate the formation in a clear and manageable way. In this example, we use two quantiles for velocity, two quantiles for acceleration, and four quantiles for angle. This simplified version is intentionally smaller than the full quantile cube described in the main text of the paper. By reducing the number of quantiles and points, we provide a concrete, easy-to-follow example that illustrates how raw data are mapped into the quantile cube, how counts are aggregated within each bin, and how proportion vectors are derived. This approach allows readers to gain intuition about the process without being overwhelmed by the complexity of a full-scale dataset, while still demonstrating all the key steps of the quantile cube methodology.

\begin{figure}[H]
\centering
\begin{minipage}{0.5\textwidth}
\begin{tabular}{c|c|c}
\textbf{Feature} & \textbf{Quantile} & \textbf{Definition} \\
\hline
Velocity ($v$) & Q1 & $\leq 3$ m/s \\
               & Q2 & $> 3$ m/s \\
\hline
Acceleration ($a$) & Q1 & $\leq 1$ m/s$^2$ \\
                   & Q2 & $> 1$ m/s$^2$ \\
\hline
Angle ($\theta$) & Q1 & $-45^{\circ}$ to $45^{\circ}$ \\
                 & Q2 & $45^{\circ}$ to $135^{\circ}$ \\
                 & Q3 & $135^{\circ}$ to $-135^{\circ}$ \\
                 & Q4 & $-135^{\circ}$ to $-45^{\circ}$
\end{tabular}
\end{minipage}%
\begin{minipage}{0.5\textwidth}
\centering
\includegraphics[width=0.8\textwidth]{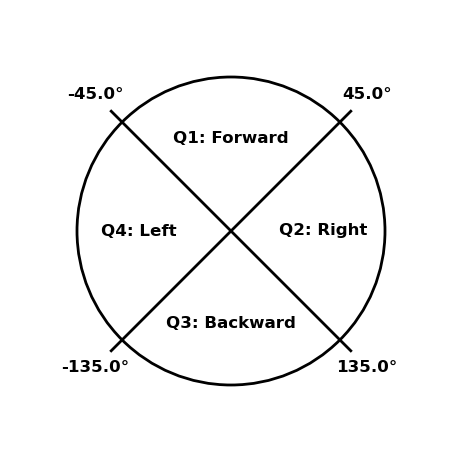}
\end{minipage}
\caption{Quantile definitions for velocity, acceleration, and angle. The table shows numeric cutoffs, while the figure illustrates the angular quantile regions.}
\label{fig:quantiles_with_angle_legend}
\end{figure}

We illustrate how five example points, listed in the first three columns of Table~\ref{table:toydata}, are mapped into this simplified quantile cube. Each point includes three features: velocity ($v$), acceleration ($a$), and angle ($\theta$). The quantile ranges, summarized in the table in Figure~\ref{fig:quantiles_with_angle_legend}, are based on a larger theoretical dataset, and the figure also provides a visual legend for the four angle quantiles. The last three columns of Table~\ref{table:toydata} provide the corresponding quantile assignment for each feature of every point.

\begin{table}[H]
\centering
\footnotesize
\begin{tabular}{c|c|c|c|c|c|c}
\textbf{Point} & \textbf{Velocity (m/s)} & \textbf{Acceleration (m/s$^2$)} & \textbf{Angle (°)} & \textbf{$v$ Quantile} & \textbf{$a$ Quantile} & \textbf{$\theta$ Quantile} \\
\hline
1 & 2.0 & 0.5 & 0 & Q1 & Q1 & Q1 \\
2 & 4.0 & 1.2 & 50 & Q2 & Q2 & Q2 \\
3 & 3.0 & 0.8 & 180 & Q1 & Q1 & Q3 \\
4 & 3.5 & 1.5 & -160 & Q2 & Q2 & Q3 \\
5 & 2.5 & 0.7 & -30 & Q1 & Q1 & Q1
\end{tabular}
\caption{Toy data points used for the quantile cube example.}
\label{table:toydata}
\end{table}

Since we now have two quantiles for velocity, two for acceleration, and four for angle, the quantile cube has $2 \times 2 \times 4 = 16$ bins. Each bin is identified by a triplet $(v_q, a_q, \theta_q)$ representing the quantile assignments for velocity, acceleration, and angle, respectively. We count the number of points in each bin and compute proportions as the count divided by the total number of data points (5):

\begin{table}[H]
\centering
\begin{tabular}{c|c|c}
\textbf{Bin $(v_q, a_q, \theta_q)$} & \textbf{Count} & \textbf{Proportion} \\
\hline
(Q1,Q1,Q1) & 2 & 0.4 \\
(Q1,Q1,Q2) & 0 & 0.0 \\
(Q1,Q1,Q3) & 1 & 0.2 \\
(Q1,Q1,Q4) & 0 & 0.0 \\
(Q1,Q2,Q1) & 0 & 0.0 \\
(Q1,Q2,Q2) & 0 & 0.0 \\
(Q1,Q2,Q3) & 0 & 0.0 \\
(Q1,Q2,Q4) & 0 & 0.0 \\
(Q2,Q1,Q1) & 0 & 0.0 \\
(Q2,Q1,Q2) & 0 & 0.0 \\
(Q2,Q1,Q3) & 0 & 0.0 \\
(Q2,Q1,Q4) & 0 & 0.0 \\
(Q2,Q2,Q1) & 0 & 0.0 \\
(Q2,Q2,Q2) & 1 & 0.2 \\
(Q2,Q2,Q3) & 1 & 0.2 \\
(Q2,Q2,Q4) & 0 & 0.0 \\
\hline
\textbf{Total} & \textbf{5} & \textbf{1.0}
\end{tabular}
\caption{Counts and proportions of points in each quantile cube bin.}
\end{table}

The final 16-dimensional representation can then be expressed as a raw count vector or as a proportion vector. The vector elements are ordered systematically: for each velocity quantile (Q1, then Q2), we cycle through acceleration quantiles (Q1, then Q2), and for each velocity-acceleration combination, we cycle through all angle quantiles (Q1, Q2, Q3, Q4). This gives us the following representations:

Raw count vector:
$$[2, 0, 1, 0, 0, 0, 0, 0, 0, 0, 0, 0, 0, 1, 1, 0]$$

Proportion vector:
$$[0.4, 0.0, 0.2, 0.0, 0.0, 0.0, 0.0, 0.0, 0.0, 0.0, 0.0, 0.0, 0.0, 0.2, 0.2, 0.0]$$

\begin{figure}[H]
\centering

\begin{subfigure}[b]{0.48\textwidth}
    \centering
    \includegraphics[width=\textwidth]{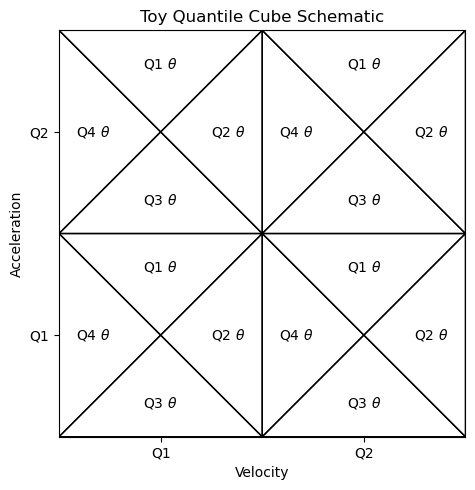}
    \caption{Schematic with labeled bins}
    \label{fig:schematic}
\end{subfigure}
\hfill
\begin{subfigure}[b]{0.48\textwidth}
    \centering
    \includegraphics[width=\textwidth]{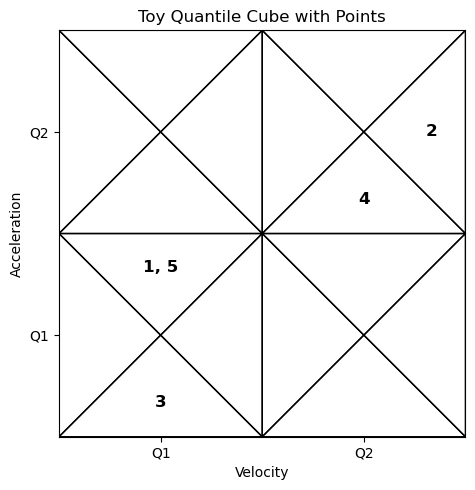}
    \caption{Point assignments to bins}
    \label{fig:numbered}
\end{subfigure}

\begin{subfigure}[b]{0.6\textwidth}
    \centering
    \includegraphics[width=\textwidth]{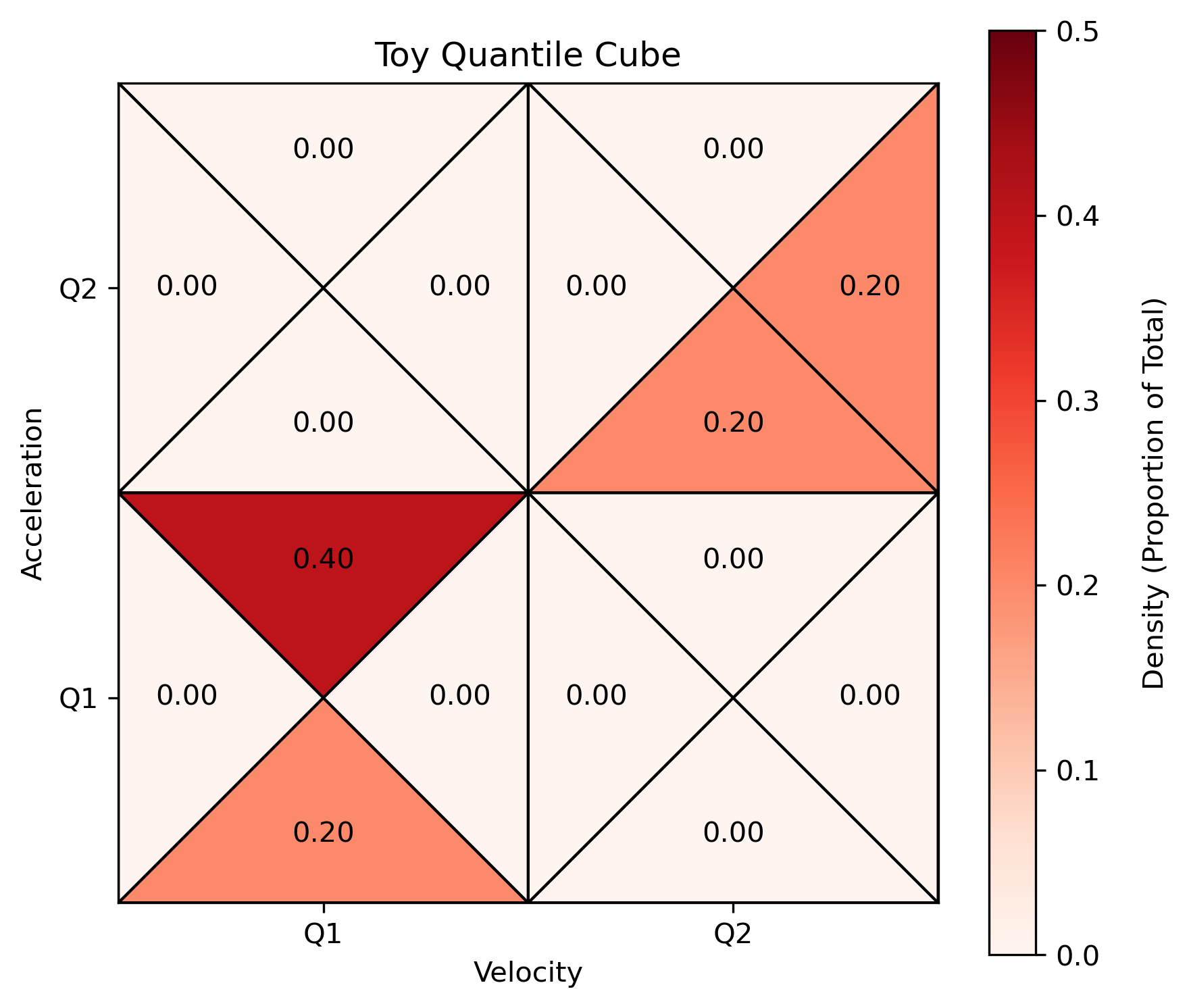}
    \caption{Final quantile cube with proportions}
    \label{fig:colored}
\end{subfigure}

\caption{Illustration of the quantile cube formation. Panel (a) shows the schematic with labeled bins, panel (b) shows the data point numbers that fall into each bin, and panel (c) shows the proportions in each bin as a colored visualization of the quantile cube.}
\label{fig:quantile_cube_all}
\end{figure}

From the proportional representation, we can then visualize the quantile cube. Figure~\ref{fig:quantile_cube_all} provides a walk-through of the creation, starting with the schematic, then into the point assignment, and concluding with the final toy quantile cube visualization.

\section{Data Preprocessing Flowchart}
\label{appendix:flowchart}
\begin{figure}[H]
    \centering
    \begin{tikzpicture}[node distance=1.5cm]
        \node (start) [startstop] {Raw \gls{gps} Data: 33 athletes, full season};
        \node (filter) [process2, below of=start] {Filter: Athletes with >25 min/half in >5 matches; Matches with >25 min in both halves};
        \node (subset) [process2, below of=filter] {Result: 198 valid athlete-match sessions (9 athletes, 23 matches)};
        \node (split) [process2, below of=subset] {Each session split into 2 halves $\Rightarrow$ 396 athlete-match-halves};
        \node (convert) [processblue, below of=split] {Convert \gls{gps} (lat/lon) $\rightarrow$ $(x,y)$ meters};
        \node (spline) [processblue, below of=convert] {Spline fit at 10 Hz; Derive velocity, acceleration, angle};
        \node (clean) [processblue, below of=spline] {Threshold low values; Apply $\log_{10}(1 + x)$ transform};
        \node (quantilebins) [process2, below of=clean] {Quantile Cube Bins Determined (5 velocity × 5 accel × 4 angle = 100 bins)};
        \node (cube) [processblue, below of=quantilebins] {Apply bins to each half $\Rightarrow$ Quantile cube per athlete-half-match};
        \node (output) [startstop, below of=cube] {Final Output: $n=396$, $d=100$ vectorized cubes};
        
        \draw [arrow2] (start) -- (filter);
        \draw [arrow2] (filter) -- (subset);
        \draw [arrow2] (subset) -- (split);
        \draw [arrow2] (split) -- (convert);
        \draw [arrow2] (convert) -- (spline);
        \draw [arrow2] (spline) -- (clean);
        \draw [arrow2] (clean) -- (quantilebins);
        \draw [arrow2] (quantilebins) -- (cube);
        \draw [arrow2] (cube) -- (output);
        
        \node (legendtitle) [below=2cm of output, align=left] {\textbf{Legend:}};
        \node [processblue, right=1cm of legendtitle, minimum width=1cm, minimum height=0.7cm] (bluebox) {};
        \node [right=0.3cm of bluebox] {Per athlete-match-half steps};
        
        \node [process2, below=0.6cm of bluebox, minimum width=1cm, minimum height=0.7cm] (redbox) {};
        \node [right=0.3cm of redbox] {Global dataset steps};
    \end{tikzpicture}
    \caption{Flowchart of data preprocessing steps from raw \gls{gps} data to quantile cube representation. Blue boxes indicate processing per athlete-match-half; light red boxes indicate global processing steps across the dataset.}
    \label{fig:preprocessing_flowchart}
\end{figure}

\end{document}